\documentclass[12pt]{article}

\usepackage{hyperref}
\usepackage{epsfig}
\usepackage{amsmath}
\usepackage{amsfonts}
\usepackage{amssymb}
\usepackage{eufrak}
\usepackage{mathrsfs}
\usepackage{graphics}
\usepackage{color}

\def\arXiv#1{\href{http://xxx.lanl.gov/hep-th/abs/#1}{#1}}

\def\be{\begin{equation}}
\def\ee{\end{equation}}
\def\ba{\begin{eqnarray}}
\def\ea{\end{eqnarray}}
\def\mc{\mathcal}

\def\appendix{{\section*{Appendix}}\let\appendix\section%
        {\setcounter{section}{0}
        \gdef\thesection{\Alph{section}}}\section}

\begin{document}


\thispagestyle{empty}
\def\thefootnote{\fnsymbol{footnote}}
\begin{flushright}
 WIS/15/06-SEPT-DPP \\
 hep-th/0609147 \\
 \end{flushright}
\vskip 0.5cm


\begin{center}\LARGE
{\bf De Sitter Quantum Mechanics\\ and Inflationary Matrix Cosmology
of\\ Self-tuned Universe}\\
\end{center}


\vskip 1.0cm

\begin{center}
\centerline{Satabhisa Dasgupta and Tathagata Dasgupta}

\medskip
\medskip
\centerline{{Department of Particle Physics,}} \centerline{{Weizmann
Institute of Science, Rehovot 76100, Israel}}

\medskip
\centerline{\tt sdasgupt, tdasgupt@wisemail.weizmann.ac.il}

\vskip 0.5cm
\end{center}

\vskip 1.0cm

\noindent We observe that the large $N$ world sheet RG in $c=1$
matrix model, formulated in \cite{ts-s,ts-ns}, with $N^2$ quantum
mechanical degrees of freedom at small compactification radius is
capable of capturing dimensional mutation. This manifests in
deforming the familiar $AdS_2$ quantum mechanics in the
minisuperspace Wheeler-de Witt (WdW) cosmology of the $2D$ quantum
gravity, obtained by the large $N$ RG with $N$ quantum mechanical
degrees of freedom only, to a modified WdW cosmology describing
tunneling to an inflationary de Sitter vacuum and its evolution. The
scale fluctuation plays an important role in providing an ansatz for
uniquely choosing the initial wave function. We observe that the
nonperturbative effects due to the $N^2$ quantum mechanical degrees
of freedom introduce explicit open string moduli dependence in the
wave function via the Hubble scale, which determines the geometry of
the true vacuum one tunnels to. The modified WdW equation also
captures controlled formation of baby universes of vanishing size
that self-tunes the de Sitter cosmological constant to be small
positive.
\date{September 2006}

\vfill \setcounter{footnote}{0}
\def\thefootnote{\arabic{footnote}}
\newpage


\tableofcontents


\section{Introduction}
\setcounter{equation}{0}

One of the central problems in quantum gravity in recent time is to
find a vacuum wave function for a consistent quantum mechanical
description of the early universe (with respect to the recent
observational data) and to choose it uniquely, if there is a
dynamical selection principle. A simple quantum cosmological
approach \cite{dw} (see \cite{v-aqc} for a review) would be to
arrive at a Schr\"odinger like (WdW) equation that would describe
tunneling of the universe from {\it nothing} to an inflationary de
Sitter vacuum \cite{v-qcis}, rather than to an Euclidean
Hartle-Hawking vacuum \cite{hh}. Once the schr\"odinger equation is
derived, one would also like to have its unique solution
\cite{v-bc}. We refer to
\cite{tryon,v-cun,v-qcu,linde-qciu-1,linde-qciu-2,zs,r-pctu,r-qmtu}
for a discussion on quantum creation of inflationary universe from
nothing. One can see recent works, for example \cite{tye-ss} and
references therein, for a comparative discussion of the two types of
wave functions in string theory context.

In string theory the universe can be chosen to be the holes in the
world sheet, that can be filled with D-branes. The dimensionality of
the universe and the issue of openness/closedness or the
homogeneity/inhomogeneity depends on how we fill these holes. Even
though these holes are world sheet quantities, they can have a good
space-time description. The one point function of these objects
describes the wave function $\psi$ of the universe with a scale
factor $l=\oint e^{\frac{1}{2}\gamma \varphi} d^2 \sigma$ ($\varphi$
is the Liouville field in a noncritical theory). In minisuperspace
approximation, the WdW equation (a BRST constraint of the field
theory)\be \frac{1}{2}(L_0+\bar L_0) | \psi\rangle=0 \label{brst}\ee
is solved for Euclidean Hartle-Hawking vacuum with homogenous FRW
cosmology. Because of reaching a huge size in a small time interval,
for most of the purposes the cosmological evolution of the
inflationary universe is semiclassical and minisuperspace
computations successfully capture a lot of features
\cite{linde-ppic}. However, in an inflationary universe, the long
wave fluctuations of the scalar field drives geometry at much larger
scales (at $l \gtrsim H^{-1} \sim M_p^{-1}$) to be highly
inhomogeneous, far away from that of a homogeneous Friedmann space.
The quantum fluctuation of the scalar field may drive some parts of
the universe to inflate eternally, whereas other parts pass the end
of inflation and undergo different phases of evolution. Much of the
important inflationary physics are actually lost in the realm of the
minisuperspace approximation, that can not get rid of an initial
presumption of global homogeneity of the universe. The initial wave
function in early universe dealing a cosmological singularity with
particle production on the horizon is essentially quantum mechanical
and is worth seeking a proper stringy description (see
\cite{linde-isc} for a review on the subject; also see \cite{zz-dmv}
for a recent study of decay of metastable vacua in $2D$ Liouville
gravity, for issues on nucleation in $2D$ Liouville gravity).

The problem can be defined in the following way. The cosmological
singularities often can be resolved by a phase of suitably chosen
closed string tachyon condensate much like the exponential wall in a
time-like Liouville theory (see \cite{silv-scst, bcraps} and
references therein). Generically the fields in such tachyon phase
become heavy and tend to evacuate configurations of real or virtual
particles or excitations that source any component of string field
(see \cite{hs} and further references). More specifically the
$S$-matrix of the system, that governs the perturbative $BRST$
consistency condition on the states in the tachyon condensate phase,
does not have perturbative asymptotic particle poles. As a result
string world sheet reaches spatial infinity in a finite world sheet
time, giving rise to holes in the world sheet. Though in special
cases $D$-brane boundary states make these holes consistent,
generically they give rise to $BRST$ anomaly that can be resolved by
considering correlations of these holes corresponding to their
unitary mapping. These correlations can be nonlocal representing
interaction between multiple world sheets \cite{abs,w-multi} and can
be complicated enough to model the dynamics in the black hole
interior (see examples discussed in \cite{hs}). Note that the above
scenario of preserving unitarity is reminiscent of the black hole
final state proposal for solving the information paradox \cite{hm}.
However, imposing cancelation of $BRST$ anomaly alone does not
specify a unique correlation and thus a unique final state, rather
yields a restricted set of states in the tachyon condensate phase.

In the light of above discussion the vacuum wave function for the
early universe can be described by a wave function for some kind of
tachyon condensate $\langle T\rangle$ that replaces the initial
cosmological singularity. A simplest undeformed condensate would
correspond to the Euclidean Hartle-Hawking vacuum whose evolution is
given by the usual WdW equation~\footnote{In time dependent
background in perturbative string theory (in type IIB and as well as
in heterotic strings) such a condensate is constructed from the
correlators supported in the weakly coupled bulk \cite{Mcgr-silv}
that seems to be a useful setup to study space-like big bang
singularities in Milne space-time. For a nonperturbative study, see
\cite{mrs}.}. However, the stringy modification to the WdW equation
can come from an anomaly in the BRST like constraint (\ref{brst})
that is resolved by a nontrivially deformed condensate $\langle
T+\delta T\rangle$ (See \cite{fst,st,hmh,bda} for other relevant
approaches/proposals of modifying the Hartle-Hawking vacua). The
allowed set of states surviving in this nontrivial deformed tachyon
condensate phase can arise from various `beyond minisuperspace'
contributions. As we discussed before, they can arise from nonlocal
correlations of the holes via multi-trace \cite{abs,w-multi} or
topology changing amplitudes that are essentially beyond
minisuperspace \cite{v-aqc}. These amplitudes describe emission or
absorption of baby universes in the WdW cosmology
\cite{scole,pol,cts1,cts2} that can self-tune the cosmological
constant to appropriate value. However, in a full nonperturbative
setup (like the one we are going to use in this paper) the deformed
condensate can contain nonperturbative objects that can feel all the
nonzero modes of dilaton. Note that, in the perturbative computation
of Euclidean vacuum in \cite{Mcgr-silv}, the tachyon condensation
effectively masses up all the closed string modes and hence the
fluctuations in dilaton too, freezing it to its bulk weakly coupled
value. Thus the setup remains perturbative there. There are other
similar evidences of perturbativity in the closed string tachyon
condensation such as repulsion felt by $D$-brane probe from winding
tachyon phase \cite{aps} (also see fate of twisted $D$-branes in
orbifolds \cite{mt,mp}).

However, in this work we will argue that the presence of
nonperturbative objects, that penetrate the tachyon wall down to
strong coupling (also see the $2D$ example in \cite{ks}), make it
possible to feel the dilaton fluctuations which play a vital role in
uniquely determining the quantum state corresponding the
nontrivially deformed tachyon condensate. We will also discuss how
the ability to feel the dilaton fluctuations can have implications
in resolving a cosmological singularity. In particular we note the
possibility of using the properties of density perturbation spectra,
such as its near flatness, to determine an ansatz for dynamical
vacuum selection via the initial profile of the scale (dilaton)
fluctuation.

In the language of WdW cosmology, computing the wave function for
the appropriate deformed tachyon condensate replacing a cosmological
singularity would require going beyond the minisuperspace by
considering general matter dependent deformations and deformations
due to higher order derivatives in scale. In other words one needs
to access the full superspace by considering the arbitrary scale and
matter fluctuations due to all the nonzero modes of dilaton and
matter. The nonzero modes of matter can be excited by appropriately
considering Neumann boundary condition on the world sheet holes.
However, the role of arbitrary matter fluctuations in a
nonperturbative tachyon condensate we are considering, is an
interesting issue that we take up elsewhere~\cite{ts-neu}. Thus
studying an initial wave function for an inflationary vacuum
eventually would be tied to a time dependent nonperturbative setup
capturing dynamical fluctuations and nonlocal processes. Even though
we start with a (sub)critical background, we believe the phenomena
we are observing is generic due to dimensional mutation brought by
these nonperturbative objects, that we explain below.

So far we focussed on the question of capturing the inflationary
nature of the initial wave function and its unique solution. Now the
issue of getting a (Lorentzian) de Sitter vacuum as opposed to an
(Euclidean) AdS is tied to the question of accessing the
supercritical regime of the theory instead of a (sub)critical one
\cite{pol,cts1,cts2}. Let us take up the example of bosonic string
theory. In $2D$ noncritical bosonic string theory (Liouville coupled
with $c=1$ matter) the minisuperspace WdW equation for $\langle
T\rangle$, both in the continuum and discrete version (free
fermionic picture of matrix quantum mechanics) is a Bessel equation
solving for an Euclidean $AdS_2$ vacuum ($\psi \sim K_{0}(\sqrt{\mu}
l)\,,~ \mu>0$) (see the reviews \cite{gm,kleb, mart} and
\cite{dcm}). The minisuperspace equation in a supercritical setup
(Liouville coupled to $c>25$) with a tachyon condensate in one of
the matter directions solves for global de Sitter in the  far future
($H_0(\sqrt{-\mu}l)$) \cite{dcm}. However, due to the simplicity of
the condensate, the vacuum wave function does not capture any kind
of inflationary scenario or proliferation of the de Sitter. The
striking feature is that, the minisuperspace WdW equation in this
case is apparently a de Sitter quantum mechanics of single matter
but with other $c=25$ matter directions fixing the Weyl anomaly.
Thus their only role is in fixing the de Sitter sign of the
cosmological constant, the Lorentzian signature of the world sheet
metric and a sensible interdependence of the scale and the string
coupling for a viable cosmological scenario. On the other hand,
remnants of $dS_2$ quantum mechanics is already there even in
minisuperspace $2D$ noncritical theory. This is evident from the
free fermion quantum mechanics, where a special class of time (or
$c=1$ matter) dependent $w_\infty$ deformations of the Fermi sea
gives rise to past and future space-like boundaries like a
space-time cosmological event horizon \cite{dk}~\footnote{See
\cite{ks, ddlm, kms} for the issue of particle production and
thermality. Also see \cite{sk} for other time dependent solutions of
similar $w_{\infty}$ deformed condensates with a cosmological
interpretation.}. Let us here remind ourselves that the
semiclassical condensate $\langle T\rangle$ giving rise to Euclidean
Hartle Hawking vacuum ($AdS_2$) of the $c=1$ theory, can be thought
of as simplest fluctuations on the Fermi surface corresponding to an
inverted harmonic oscillator potential \cite{gm}. Thus perhaps in
the nonperturbative setup, the nontrivial time dependent
deformations of this Euclidean Hartle Hawking vacuum in the
(sub)critical theory is carrying out some kind of dimensional
mutation by capturing the physics of a supercritical regime (see
discussions in \cite{silv-dm} and references therein for comments on
dimensional mutation in closed string tachyon condensation in the
context of cosmological singularities).

In this paper we put together all the ingredients of the problem in
the context of the large $N$ RG \cite{ts-s,ts-ns}, a Wilsonian world
sheet RG, of Dirichlet boundaries in $2D$ noncritical string theory
(the FZZT branes \cite{fzzt} with Dirichlet boundary condition on
matter). They describe the quantum cosmology of the one dimensional
universes with homogeneous matter. The wave function of the universe
is given by the one point function of the loop operator
$W(l,t)=\frac{1}{N}\mbox{Tr}\exp[l \Phi(t)]$ \cite{bdss, m, ms,
mss}. Note that the loop operators we study also have space-time
interpretation. Here we compute the wave function by performing the
large $N$ RG of the one point function of the loop operator of the
noncritical theory around a $c=1$ fixed point at a string scale
compactification radius (a self-dual radius $R \sim 1$). In fact, in
our RG scheme, the Dirichlet boundary condition on matter at a $c=1$
fixed point prefers a small compactification radius. As a result the
large $N$ RG deals with all the $N^2$ quantum mechanical degrees of
freedom, instead of the $N$ eigenvalues only as in minisuperspace
free fermionic description. This translates into unavoidable large
fluctuation of dilaton and nonlocal effects associated to the
presence of the off diagonal elements (that can not be integrated
out in $c=1$ theory on a small circle) and presumably release of
large number of nonperturbative states \cite{gk1,gk2,bk} causing a
dimensional mutation. Thus the RG captures a nontrivially deformed
condensate that modifies the WdW equation by scale derivatives of
all orders, a nontrivial time dependence and nonlocal deformations
corresponding to emission and absorption of point-like baby
universes. Strikingly the process of emission and absorption of
babies are controlled by $O(1/N)$. The short distance wave function
of this modified WdW gives a tunneling wave function that
corresponds to birth or nucleation of the universe into an
inflationary de Sitter vacuum. Here let us briefly mention that in
the recent semiclassical study of the decay of weakly metastable
vacua in $2D$ Liouville gravity in \cite{zz-dmv}, gravitational
inflation inside the region of lower energy phase due to droplet
fluctuation in the critical swelling process modifies the standard
exponential suppression of the nucleation probability to a power law
one, much like the behavior of the dynamical lattice Ising model.
Now the far future wave function gives a global de Sitter as would
have been given by a $dS$ minisuperspace theory. The scale
derivatives are playing an important role. Truncating them to
quadratic power gives only the global de Sitter
($J_0(\sqrt{-\mu}l)$) in the far past like a de Sitter
minisuperspace theory instead of a tunneling wave function. The
arbitrary scale fluctuations help to fix the initial condition
uniquely.

It is interesting to observe that, the time dependent
nonperturbative deformations in noncritical theory is doing
something nontrivial to enhance the effective space-time dimensions
and hence accesses the Lorentzian de Sitter geometry dwelling in a
supercritical regime. Presumably the dimensional mutation is due to
large number of nonperturbative states belonging to the nonsinglet
sector of the matrix quantum mechanics appearing at the self-dual
radius. The real nature of these states (whether they are best
described as world sheet vortices or long winding strings) is yet to
be understood \cite{gk1,gk2,bk,malda-ls} (also see \cite{ts-ns, kkk}
for a discussion on the nature of these states in the context of
$2D$ black hole). It has been pointed out in \cite{silv-dm} that in
the case of spacelike singularities studied in \cite{Mcgr-silv} and
in similar cases with positively curved spatial slices, the
singularity is lifted by a tachyon phase with fewer degrees of
freedom, as opposed to the cases corresponding to negatively curved
spaces. This is a candidate big bang case where with forward time
evolution along the larger direction of the compactification circle
the dimensionality ($c_{eff}$) increases. This is similar to the
realistic inflationary cosmology where universe evolves in the
direction of the evolution of de Sitter entropy
\cite{s-dscft,s-infl-dscft}). Note that, in our case the universe
evolution is not in the direction of decreasing $R$, instead is in
the direction of increasing $l$. The big bang singularity in our
picture sits at small scale ($l \sim H^{-1}$) and is resolved by
popping up of a simple tachyon condensate (Euclidean vacuum) at
large $R$. Whereas tuning the radius to string scale deforms the
condensate setting in dimensional mutation to supercritical regime.
In this case an inflationary de Sitter vacuum pops up at $l \sim
H^{-1}$ to resolve the singularity. Nevertheless the universe
evolution in our picture does take place in the direction of
evolving de Sitter entropy. This is because for $l<H^{-1}$ we have
tunneling to the corresponding vacuum from nothing. Thus in effect
the $R$ direction maps to the set of initial condition in string
theory that maps to the available set of vacuum ($AdS$ or $dS$)
through the large $N$ RG (see the figure \ref{fchart-intro} below).
However, as we will see, the dynamical vacuum selection here depends
on other factors (dilaton fluctuation) that uniquely specify the
initial tunneling wave function. As pointed out in \cite{polyakov},
the universe we live in seems to have some mechanism of self tuning.
Perhaps the large $N$ RG incorporating dilaton fluctuation and
nonlocal effects due to multiple world sheets inherently exhibits
such self tuning that enables one to flow to a more realistic
vacuum, starting around a $c=1$ fixed point. It will be interesting
to generalize the problem, in particular for an inhomogeneous
universe with arbitrary scale and matter fluctuation, which we will
report later \cite{ts-neu}.

\begin{figure} [ht]
\centering
\input{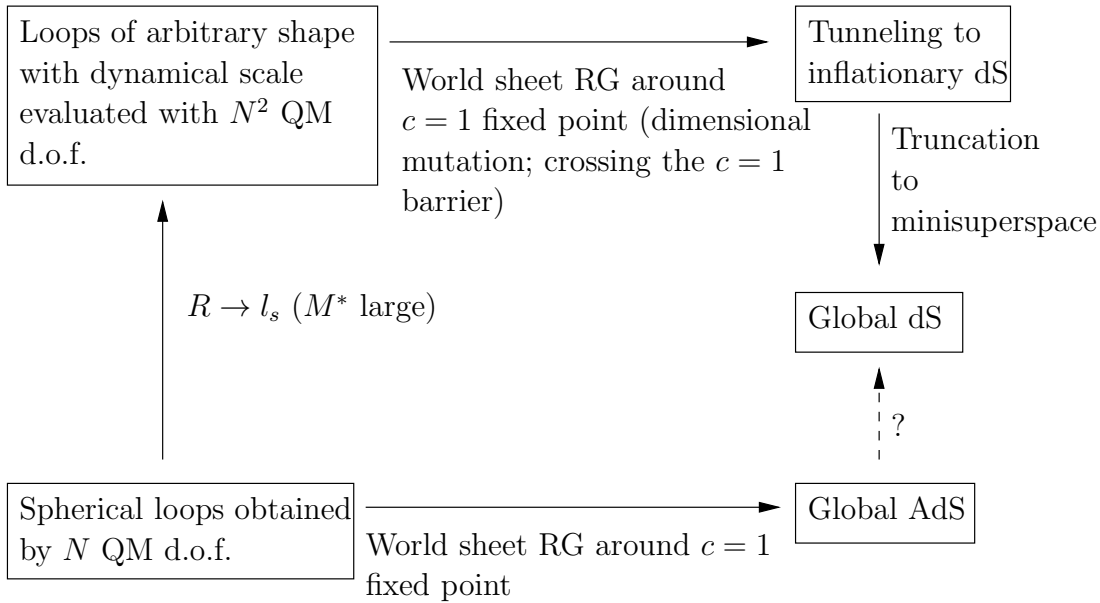} \caption{The mapping of the
set of initial conditions in string theory to the set of available
vacua. The broken arrow denotes possible time dependent deformations
of the Fermi surface picture that may connect the two sets of
vacua.} \label{fchart-intro}
\end{figure}

The plan of the paper is as follows. In section 2 we review and
elaborate the problem of inflationary de Sitter in $2D$ cosmology
and summarize our results and critical observations. In section 3 we
review our previous work~\cite{ts-s} on large $N$ RG analysis in
$c=1$ matrix model. In section 4 we deduce the familiar
minisuperspace WdW equation of free fermions, namely the $AdS_2$
quantum mechanics, from large $N$ RG. In section 5 we compute its
deformation, the generalized WdW equation for one dimensional
homogenous universe with arbitrary shape and scale fluctuations. In
section 6 we identify the Hubble scale and then solve for the wave
function at small and large scale limits and discuss the
cosmological implications: tunneling to inflationary de Sitter in
the far past and an expanding global de Sitter in the far future. In
section 7 we discuss the topology changing amplitudes captured by
large $N$ RG and suppression of baby universes and self-tuning of
cosmological constant.


\section{$2D$ cosmology and inflationary de Sitter vacuum}
\setcounter{equation}{0}

Let us now review and elaborate on the basic philosophy of the
problem in bosonic string theory.


\subsection{The minisuperspace WdW cosmology in (sub)critical
and supercritical theory}

Let us consider Euclidean quantum gravity described by Liouville
field theory coupled to $c=1$ matter or its discrete matrix quantum
mechanics version, where one has a good nonperturbative description
of the time dependent physics. The tachyon field $T(\varphi,
X(\sigma))$ on the world sheet is directly related to the
macroscopic loop fields $W(l,t)$ \cite{gm,kleb,mart,bdss,m,ms,mss}
via

\be W(l,t)\sim e^{-\frac{1}{2} Q \varphi}~ T(\varphi,t)\,.\ee Here,
$\varphi$ is Liouville field and $Q$ is the slope of the linear
dilaton background ($Q=\sqrt{2}$ for $c=1$ theory). The relation can
be intuitively understood by considering the first order fluctuation
$\delta W$ that corresponds exactly to the tachyon wave function and
satisfies minisuperspace WdW equation up to a factor of coupling
constant. The vacuum expectation $\langle W(l,t)\rangle$ is given by
the Hartle-Hawking type wave function $\psi(l)$. In the matrix
quantum mechanics, these are nothing but the $FZZT$ branes or the
one dimensional boundaries of length $l$

\be \psi(l)=\frac{1}{N} \langle \mbox{Tr} e^{l \Phi(t)}\rangle
\,,\ee with Dirichlet boundary condition on the world sheet matter

\be W(l,t)_{cont}~\sim~\delta(\oint e^{\frac{1}{2}\gamma
\varphi}-l)~ \delta (X(\sigma)-t)\,. \label{dbc}\ee Here
$W(l,t)_{cont}$ represents the loop operator in the corresponding
continuum Liouville theory. Their wave function, computed by the
eigenvalue representation or the free fermion representation of
matrix quantum mechanics \cite{m,ms,mss}, satisfies the same
minisuperspace WdW equation as in the continuum Liouville field
theory (accompanied by proper identification of the matrix model
couplings with the continuum parameters)

\be \Big[-(l_0 \frac{\partial}{\partial l_0})^2+4 \mu l_0^2+\nu^2
\Big] \psi(l_0)_{h=0}=0\,,~~~~l_0=e^{\frac{\gamma}{2}
\varphi_0}\,.\label{minisupwdw} \ee In matrix model computation, the
all genus version of (\ref{minisupwdw}) for zero Fourier mode of
matter modifies the WdW by adding a $-g_s^2 l_0^4$ term in the
potential. Note that, the free fermion computation we mentioned here
is computed in minisuperspace ({\it i.e.} $\varphi$ space is
truncated to $\varphi_0$). This is manifest in the massless scalar
(the fluctuation of the eigenvalue density) quadratic action that
exactly maps to the minisuperspace WdW action of the Liouville field
in a suitable coordinate in the space of eigenvalues \cite{gm}. Both
the continuum and the matrix model solutions for the minisuperspace
Hartle-Hawking vacuum (an undeformed tachyon condensate) are just
Bessel functions with simple FRW cosmology. The solution decaying at
large length is the non-normalizable wave function
$K_{\nu}(2\sqrt{-\mu}l_0)$. In fact $K_{0}(2\sqrt{-\mu}l_0)$
describes AdS vacuum ($\mu>0$) that has simple big bang and big
crunch \cite{dcm}. Thus the free fermion quantum mechanics
(\ref{minisupwdw}) with $\mu>0$ is a quantum mechanics for $AdS_2$.
From the point of view of loop operator, it describes dynamics of
spherical loops with a homogeneous matter. Note that a wave function
given by Bessel $J_0(2\sqrt{-\mu}l)$ and Hankel
$H_0({2\sqrt{-\mu}}l)$ would have described the behavior of global
de Sitter in the far past and far future respectively \cite{dcm}, as
it will arise in a different context. However, as we will see, these
Bessel type of solutions describe merely homogeneous FRW geometries
and do not seem to capture any inhomogeneity introduced by
inflationary scenario with cosmological time evolution.

In terms of the free fermion phase space, the undeformed tachyon
condensate can be described by the physics of a static Fermi sea
corresponding to the eigenvalue hamiltonian with inverted oscillator
potential filled up to a level $\mu$. The tachyon wave function can
be visualized as the fluctuations $\delta W$ along the Fermi surface
$H(p,\lambda)=\mu$. Now a general potential $V(\lambda)$ would
spatially deform (fold, shift or open) the Fermi surface
$H(p,\lambda)=p^2-\lambda^2=\mu$ to a Fermi surface of an arbitrary
shape $H(p,\lambda)=p^2+\frac{1}{2} V(\lambda)=\mu$. This introduces
higher derivatives in the minisuperspace WdW equation that modifies
the genus zero tachyon wave function according to \be \Big[
\partial_t^2+l_0^2 V\big(\partial/\partial l_0 \big)+\frac{l_0}{2}
V' \big(\partial/\partial l_0 \big) \Big]~\delta W_{h=0}=2 \mu
l_0^2~\delta W_{h=0}\,. \label{higher-derivative}\ee Explicit time
(matter) dependent deformation (tearing into droplet, opening and
draining to the other side) of the Fermi surface result in
nontrivial time dependent coefficients in the kinetic and potential
terms of the WdW equation. Effectively it boils down to a time
dependent $\mu(t)$ and a time dependent over all multiplication
factor in the tachyon wave function. A  class of integrable
($w_{\infty}$) time dependent deformation have been shown to have
cosmological interpretations (e.g. bounce cosmology) \cite{sk}.

For inflationary scenario, one needs an appropriate temporal and
spatial deformation of the condensate such that the modified WdW
gives rise to some kind of nucleation to a de Sitter vacuum at UV
(or in the far past) and its proliferation. Note that from
(\ref{minisupwdw}), a naive continuation to the de Sitter geometry
($\mu <0$) by dialing the sign of cosmological constant $\mu$ is
subtle and apparently does not make sense for the eigenvalue
dynamics of the $c=1$ matter. There the $\mu \to 0$ corresponds to
instability due to spilling of the eigenvalues from the top of the
inverted oscillator potential (the continuum limit in the double
scaling procedure) and the $\mu <0$ side is unbounded. However,
considering an interesting class of $w_{\infty}$ deformations
corresponding to special class of time dependent perturbations of
the Fermi sea (opening and draining of the Fermi sea to the other
side), one can show formation of spacelike horizons either or both
at past and future spacelike infinities \cite{dk} and also particle
production \cite{ddlm,ks,kms}. Strikingly these properties indicate
remnant of some kind of de Sitter quantum mechanics. Thus it is not
meaningless in $c=1$ matrix model to look for most general
(integrable and nonintegrable~\footnote{All the integrable
$w_{\infty}$ deformations can be written in minisuperspace variable
$\varphi_0$ whereas the most general deformations in (accessible)
superspace are all not necessarily integrable.}) time dependent
deformation of the tachyon condensate that will give rise to wave
function exhibiting nucleation to de Sitter vacuum and possibly its
proliferation too. For $c \ge 25$ matter a global de Sitter geometry
at large scale can be probed by considering a Liouville-Sine-Gordon
model of a inhomogeneous tachyon condensate \cite{dcm}. The
condensate is in the form of gravitationally dressed cosine matter
potential in one of the matter (spatial) direction and is homogenous
in remaining scalar fields
 that only play a role through the Weyl
anomaly (they are all massless and remain in their initial state
throughout the evolution of the universe). The minisuperspace WdW
equation is then modified by a matter dependent cosine term in the
potential. This gives a solution exhibiting global de Sitter
geometry at large scale though it recollapses at small scale.
However, such a condensate is still too simple to catch
proliferation of the de Sitter or particle production.

In this respect one has to keep in mind that the quantum cosmology
for $c=1$ matter has few subtleties that are better dealt in $c \ge
25$ regime \cite{cts1,cts2,pol}. This is because the length scale of
the universe behaves as $l=\oint d^2 \sigma~e^{\gamma \varphi}\,,$
where $d^2 \sigma e^{\gamma \varphi}$ is conformally invariant ({\it
i.e.} $Q=\gamma+2/\gamma$). For $c=1$, $Q$ and hence $\gamma$ are
real and the length scale grows proportionally with the string
coupling $g_s \sim e^{\frac{1}{2} Q \varphi}$. The universe is then
small at weak coupling and large at strong coupling, which is
unrealistic for doing cosmology. For $c\ge 25$, $Q$ and $\gamma$ are
purely imaginary. Thus one Wick rotates $\varphi \to -i \varphi$
getting a length scale inversely varying with string coupling $g_s
\sim e^{-\frac{1}{2}Q \varphi}$. The wick rotation has an added
advantage that it makes the kinetic energy of the time-like
Liouville coordinate negative (analogous to that in $4d$ Euclidean
gravity) so that the world sheet geometry has a Lorentzian
interpretation. Thus for technical reasons we discussed so far, $c
\ge 25$ matter has been preferred over the $c=1$ matter theory in
probing de Sitter cosmology on the world-sheet~\cite{dcm}. Finally
for  $c=1$ theory at strong coupling, formation of baby universes of
macroscopic size is favored that may absorb the inflating core of
the universe and force it to recollapse, once the fluctuation drives
the inflaton off its potential maxima. In single closed universe
cosmology of gravity coupled to $c=1$ matter in $\mc{R} \times S^1$,
where the winding strings in a Sine-Gordon Liouville setup can
provide topological defect acting as a seed, such a mechanism
disallows possible eternal topological inflation.


\subsection{The modified WdW cosmology: results and observations}

In this paper, we use the general framework of large $N$ RG approach
in matrix quantum mechanics on a circle \cite{ts-s,ts-ns} that works
with $N^2$ quantum mechanical degrees of freedom instead of $N$
eigenvalues and thus probes physics beyond the free fermionic
approximation. The dilaton emerges in a more complicated way
involving all the off-diagonal degrees of freedom and hence
$\varphi$ is not simply truncated to $\varphi_0$ but is dynamical in
all the modes. Using this approach we derive a generalized WdW
constraint and the corresponding (all genus) wave function for FZZT
branes with Dirichlet boundary condition on matter. Thus the quantum
cosmology we study is that of one dimensional closed universe of
arbitrary shape $l=\oint d^2\sigma e^{\sqrt{2} \varphi}$ with
homogeneous matter $t=\hat x$ and arbitrary scale fluctuation
$\delta l \sim \oint d^2 \sigma e^{\sqrt{2} \varphi}\sqrt{2} \delta
\varphi$.

In matrix quantum mechanics picture the universe of length $l$ is
created by the macroscopic loop operator
$\frac{1}{l}\mbox{Tr}\phi_N^l(t)$ that cuts a hole of lattice length
$l$ on the discretized string world sheet. They have the same
continuum limit as the lattice loop operator
$W(l,t)=\frac{1}{N}\mbox{Tr}\exp(l~\phi_N(t)/a)$, where $\phi_N(t)$
are $N \times N$ hermitian matrices and $a$ is a lattice spacing,
and can be studied in this form. The commonly used boundary
condition on these one-dimensional universes is the one that
constrains the boundary length to be $l$ throughout which the matter
field takes a single value (say $t=\hat x$). It will be more useful
to compute the wave function for the one point function of the loop
operator with one puncture or the wave function for a boundary with
one marked point

\be \Psi(l)=\frac{\psi(l)}{l} \,,~~~~~~\psi(l)=\langle W(l,\hat
x)\rangle = \langle N^{-1}~\mbox{Tr}\exp [l \phi(\hat x)]\rangle
\label{opfop}\,.\ee A large length fluctuation over an arbitrary
shape of loops is the dominant effect at small scale and hence is
crucial to explain the inflationary scenario of early universe.
Below we summarize our results and critical observations.

\vskip 0.5cm
\begin{flushleft}
{\it A nontrivial condensate}
\end{flushleft}

The world sheet RG modifies the WdW constraint by adding derivatives
with respect to scale of all degrees and introducing nontrivial
matter dependence. Along the line of our discussions above, this
amounts to capturing physics of a nontrivial and complicated
(perhaps nonintegrable) spatially deformed time dependent tachyon
condensate. The inflaton field or (in this case) the matter field
acquires a light (exponentially vanishing) mass and thus have no
problem in executing a slow roll. The world sheet RG also adds a
$O(1/N)$ correction that has inherently quantum mechanical origin
coming from the anomaly in the Callan-Symanzik equation.

\vskip 0.5cm
\begin{flushleft}
{\it The quantum creation of inflationary de Sitter vacuum}
\end{flushleft}

The short distance $ l < H^{-1}  \sim M_p^{-1}$ wave function, which
we obtain, is a tunneling wave function describing quantum creation
of a closed universe from {\it nothing} and its nucleation to an
inflationary de Sitter vacuum

\be \Psi(l<H^{-1}) \sim \exp[-c H^2 l^2]\,.\label{twf}\ee At large
scale $l>H^{-1}$ the evolution of the universe is classical. Thus at
large scale the wave function reproduces the semiclassical behavior
of an expanding global de Sitter geometry, in spite of higher
derivative and other quantum effects. The relevance of the higher
derivatives at short distance can be realized from the fact that, in
this domain a quadratic WdW will give rise to a far past global de
Sitter ($J_0(\sqrt{-\mu}l)$, $\mu<0$) only, instead of a tunneling
wave function.

\vskip 0.5cm
\begin{flushleft}
{\it Suppression of baby universes}
\end{flushleft}

The baby universes generated ({\it i.e.} emitted or absorbed by loop
splitting and reconnection) in this cosmology are of vanishing sizes
and their formation is suppressed by higher orders in $1/N$. Their
amplitude comes from nonlinear part of the WdW equation, generated
by multi-trace contributions in the matrix quantum mechanics loop
dynamics, that can be represented schematically by a complicated
`$*$' operation \be \mc{H}\psi+\psi * \psi = 0\,,\ee where $\mc{H}$
is the modified WdW operator as captured by the world sheet RG. The
one point function $\psi$ is a Hartle-Hawking wave function in the
sense of \cite{gm}. The amplitude corresponding to the `$*$'
operation is coming from some anomaly, breaking the scale invariance
imposed by a linear WdW constraint $\mc{H} \psi=0$, originating from
nonlocal contribution from multiple world sheets. The vanishing size
of the baby universes and suppression of their amplitude helps in
continuation of inflation once it onsets. Note that, the birth of
baby universes involves topology changing and thus inherently is a
beyond minisuperspace process. Such amplitudes get contributions
from the points in the interior of different topological sectors of
the superspace \cite{v-aqc}. We observe that the sub-leading
contribution from the emission of a single baby universe self-tunes
the de Sitter entropy to be small positive.

\vskip 0.5cm
\begin{flushleft}
{\it Emergence of Hubble scale}
\end{flushleft}

In $2D$ the gravitational coupling is dimensionless and there is no
proper Planck scale. The Planck scale assumed to be spontaneously
induced at $g_s \sim 1$.  However, in the large $N$ world sheet RG,
the Planck scale emerges naturally as the Hubble scale for the
problem. The potential in the generalized WdW equation for the one
point function of loop operator with a marked point helps to
identify the required inverse Hubble scale $H^{-1}$, which is very
small for $c=1$ case. This would be favorable for quantum tunneling
of universe from nothing through an inflationary scenario. The
potential below this scale has a tunneling type barrier. The
geometry at $H^{-1}$, at which the universe pops out from the
classically forbidden region, is a de Sitter geometry.

\vskip 0.5cm
\begin{flushleft}
{\it The open string moduli dependence and the geometry of the true
vacuum}
\end{flushleft}

In our approach, the nonperturbative effects due to the $N^2$
quantum mechanical degrees of freedom is observed to be introducing
explicit open string moduli dependence in the wave
function\footnote{We thank Micha Berkooz for a discussion on this
issue.} and in the Hubble scale. Such a dependence is not present in
the wave function when we explicitly work with $N$ quantum
mechanical degrees of freedom. The open string moduli dependence
does not directly affect the inflationary scenario, as the tunneling
wave function solution exists for any sensible constant $H$.
However, the explicit dependence of the Hubble constant on the open
string moduli determines the geometry of the true vacuum one tunnels
to. It will be extremely interesting to investigate whether the
world sheet vortices are responsible for some kind of spontaneous
symmetry breaking that chooses the ground state.

\vskip 0.5cm
\begin{flushleft}
{\it Choice of initial condition}
\end{flushleft}

A problem of describing creation of universe, in particular an
inflationary universe, always has subtleties with choice of unique
initial condition corresponding to `nothing' \cite{tye-ss}. This has
long been a subject of big debate (see for example \cite{v-qcis}).
With introduction of more degrees of freedom as we go beyond
minisuperspace and derivatives of all degrees, the situation is even
worse. One would encounter an infinite dimensional initial condition
(for example, specifying initial value of all the derivatives of the
wave function) in all the degrees of freedom involved ($l_0, \delta
l$). However we observe that a gaussian profile of initial scale
fluctuation $\psi(\delta l) \sim e^{-\delta l^2}$ provides all the
required initial conditions and thus uniquely determines a tunneling
wave function at UV. In other words, the wave function $\psi(\delta
l)$ for initial scale fluctuation can be viewed as a co-moving
(gravitational) wave detector to capture the inflationary vacuum.

\vskip 0.5cm
\begin{flushleft}
{\it The arbitrary scale fluctuations}
\end{flushleft}

The arbitrary scale fluctuations, taking us beyond the
minisuperspace, are playing a pivotal role in defining a proper
cosmological picture here. Because of the scale fluctuation, the
actual volume of the one dimensional universe can no longer be
definitely small at weak coupling and large at strong string
coupling. A large fluctuation of the form $\delta \varphi \sim
e^{-\sqrt{2}\varphi}-e^{+\sqrt{2}\varphi}$ would drive the scenario
to the opposite. Now the length scale of the universe $l=\oint
e^{\varphi_0+e^{-\sqrt{2}\varphi}-e^{+\sqrt{2}\varphi}}$~ becomes
small as $\varphi \to \infty$ (strong coupling) and grows as
$\varphi \to -\infty$ (weak coupling).

Amusingly, such a fluctuation in $\varphi$ would imply a scale
dependence of the power spectrum of primordial density fluctuation
for our one dimensional universe (naively, the density perturbation
$\frac{\delta{\rho}}{\rho} \sim -\delta \phi \sim
e^{+\sqrt{2}\varphi}\sim l$). Strikingly, the tunneling wave
function fixed by the initial gaussian scale fluctuation profile
captures a vanishingly small power-like scale dependence (like a
real-world nearly non flat CMB anisotropy). Let us define the
probability of finding the universe between a scale $l$ and
$l+\delta l$ as $l \gtrsim H^{-1}$ at a time $0+$ after the big bang
to be $\frac{\delta l}{l}= P(l \to l+\delta l)=\frac{\psi(l+\delta
l)}{\psi(l)}\sim \Big(\frac{H^{-1}}{l}\Big)^2$. For a small universe
($l<1$) and for an estimate $l \sim H^{-0.99}$ to realize the bound
$l \gtrsim H^{-1}$, the power spectrum is slightly red-shifted \be
\frac{\delta \rho}{\rho} \sim l^{0.02}\,,~~n_s \sim 0.96\,. \ee

\vskip 0.5cm
\begin{flushleft}
{\it Arbitrary matter fluctuations}
\end{flushleft}

An arbitrary matter fluctuation is crucial to capture the particle
production and also the formation of large scale inhomogeneities in
inflation (as seen by comoving particle detectors). In matrix
quantum mechanics this is a more delicate issue as it would require
to consider wave function for a $D1$ brane with Neumann boundary
condition on both Liouville and matter coordinates. In some sense
this implies dominance of the kinetic term in the matrix quantum
mechanics action (hence presumably the dominance of the nonsinglet
sector), over the potential term, that transforms as singlet under
$SU(N)$ rotation. The angular (off-diagonal) degrees of freedom in
the kinetic term can not be integrated out except for large
compactification radius or for matrix quantum mechanics on a line
\cite{gk1,gk2,bk}.  In other words, it would need a full open string
description of matrix quantum mechanics. In \cite{minahan-open} the
problem have been dealt to some extent by considering interacting
spin chains. In an upcoming paper \cite{ts-neu} we will extend the
case studied here to an inhomogeneous universe (the Neumann boundary
condition on matter) to study arbitrary matter fluctuations. The
case with Neumann boundary conditions in both matter and Liouville
directions has not yet been solved in matrix quantum mechanics.

\vskip 0.5cm
\begin{flushleft}
{\it The factor ordering ambiguity}
\end{flushleft}

The arbitrary scale fluctuation allows a small universe at strong
coupling. As $l \to 0$, the potential in the WdW equation $V(l,\hat
x) \to 0$ and hence the initial wave function could be incoming or
outgoing plane waves (FRW geometries). However, large $N$ RG
captures a factor ordering ambiguity $p \ne 1$ of the quadratic
differential operator $(l^p\frac{\partial}{\partial
l}l^p\frac{\partial}{\partial l})$ in the WdW equation. This bars
the wave function from being only plane wave at small scale, thus
encouraging an inflationary scenario.

\vskip 0.5cm
\begin{flushleft}
{\it Instability and third quantization}
\end{flushleft}

Intuitively one can imagine for eternal inflation, in which the
space-time of universe nucleates in a fractal structure, a violent
fluctuation in all modes of $\varphi$ would be necessary. On the
other hand, such extra dynamical fields beyond minisuperspace
usually give rise to unstable scalars and hence negative norm states
leading to recollapse of the universe \cite{v-qcis}. However, in our
approach, the situation is quite different. The theory of loops on
world sheet can be seen as some kind of third quantization (see for
example \cite{gm,ms} and references therein). Then the problem of
negative norm state is not relevant as the wave function of the
universe is actually a quantum field operator (the loop operator
itself) composed of creation and annihilation operators for the
universe and the WdW equation follows from the third quantized
action. The generalized WdW operator in large $N$ RG emerges as a
natural gauge condition for loops of arbitrary shape with arbitrary
scale fluctuations (from the anomaly in the Callan-Symanzik equation
for the loop operator), that takes care of negative norm states of
the dynamical fields.

\vskip 0.5cm
\begin{flushleft}
{\it The minisuperspace limit}
\end{flushleft}

To discuss the validity of our derivations we reproduce the
minisuperspace $AdS_2$ quantum mechanics by large $N$ RG on the
space of matrix eigenvalues. It is very interesting to note that the
world sheet RG somehow captures a self-tuned deformation of the
$AdS_2$ quantum mechanics of the $N$ degrees of freedom (the matrix
eigenvalues) to that of a $dS_2$ quantum mechanics of all the $N^2$
degrees of freedom. Perhaps it is an indication that there is such a
self-tuning mechanism of dynamical vacuum selection in real world
that we will be able to capture by going beyond the minisuperspace
approximation.

\vskip 0.5cm
\begin{flushleft}
{\it D-instantonic picture for tunneling?}
\end{flushleft}

It will be certainly interesting to understand if there is any
instantonic picture for the tunneling. In free fermionic
representation, the vacuum with de Sitter sign of the cosmological
constant ($\mu <0$) has an instability due to an instantonic
tunneling amplitude $O(e^{-N})$ from the other (AdS) side and is
central to the nonperturbative formulation of the theory
\cite{shenker}. The Liouville theory coupled to $c=1$ matter does
have D-instantons (ZZ branes) of $c=1$ noncritical string theory
localized in the strong coupling region ($\varphi \to 0$) \cite{zz}.
It has been interpreted that the FZZT brane (extended in $\varphi
\to -\infty$) probes the $\varphi$ direction up to a certain scale
(until $g_s \sim 1$ analogous to  $l \sim H^{-1}$ in our cosmology),
beyond which it dissolves. Below this scale the open string degrees
of freedom are only these D-instantons stuck at $\varphi \to \infty$
which can be thought of as the remnant from the superposition of a
FZZT brane stretching from weak to strong coupling and then back to
weak coupling region \cite{mart-ar}. Thus the wave function for FZZT
branes at small scale $l<H^{-1}$ actually computes a D-instanton
tunneling amplitude in all genus. Note that, unlike the tunneling
event in $N$ degrees of freedom picture (eigenvalue representation),
here the tunneling event captured by the dynamics of $N^2$ degrees
of freedom ends in a stable de Sitter vacuum.

\vskip 0.5cm
\begin{flushleft}
{\it An artifact of the formalism?}
\end{flushleft}

One might wonder whether all these nice aspects captured by large
$N$ RG is in some sense an artifact of the formalism itself or they
are really happening. To this end, we need to clarify whether the
small matrix field (small $\Phi_N(t)$) approximation, which is the
only approximation in the large $N$ RG, is in contradiction with the
large eigenvalue at the strong coupling (eigenvalue D-instanton) or
at the point of spilling~\footnote{We thank Emil Martinec for
pointing this out.}, or more generally to a large quantum
fluctuation in the matrix field variable itself. However, in
integrating out each momentum shell ($v_{1 \times N}\,, v^*_{N
\times 1}\,, \lambda_{N+1}=0$) from a $(N+1) \times (N+1)$ matrix
$\Phi_{N+1}(t)$, the remaining $\Phi_N(t)$ part that enters in the
interaction vertex $g~\Phi_N$ is never approximated to be small in
an absolute sense. Rather it is assumed lighter compared to the
integrated out degrees of freedom which become heavier by a
rescaling $v \to \tilde v=\sqrt{N} v$. This makes the interaction
vertex lighter compared to external integrated out momenta, thus
facilitating in the expansion of its exponential to get Feynman
diagrams.

\vskip 0.5cm
\begin{flushleft}
{\it World sheet vs space-time observer?}
\end{flushleft}

Our study of the cosmology is performed with respect to the world
sheet observer, which is meaningful as the observer should be a part
of the universe. However, even though the wave function for the
universe we are describing is computed on the world sheet, the loop
operators do have a proper space time interpretation and the
cosmology could be translated on that language.

\vskip 0.5cm
\begin{flushleft}
{\it Euclidean vs Lorentzian picture?}
\end{flushleft}

It appears that the world sheet large $N$ RG with $N^2$ quantum
mechanical degrees of freedom automatically captures the Lorentzian
de Sitter vacuum of the supercritical theory instead of the
Euclidean $AdS$ vacuum of the subcritical theory, the later being
captured by the large $N$ RG of $N$ quantum mechanical degrees of
freedom. Perhaps the world sheet RG has some nontrivial dynamics
that leads to a growth of space-time dimensions\footnote{We thank
Eva Silverstein for a discussion on this.}.

In figure \ref{fchart-ch2} below, we summarize the main results in a
schematic form. The broken arrow represents a possible $w_\infty$
deformations showing past and future spacelike infinities as was
also included in figure \ref{fchart-intro}.

\begin{figure} [ht]
\centering
\input{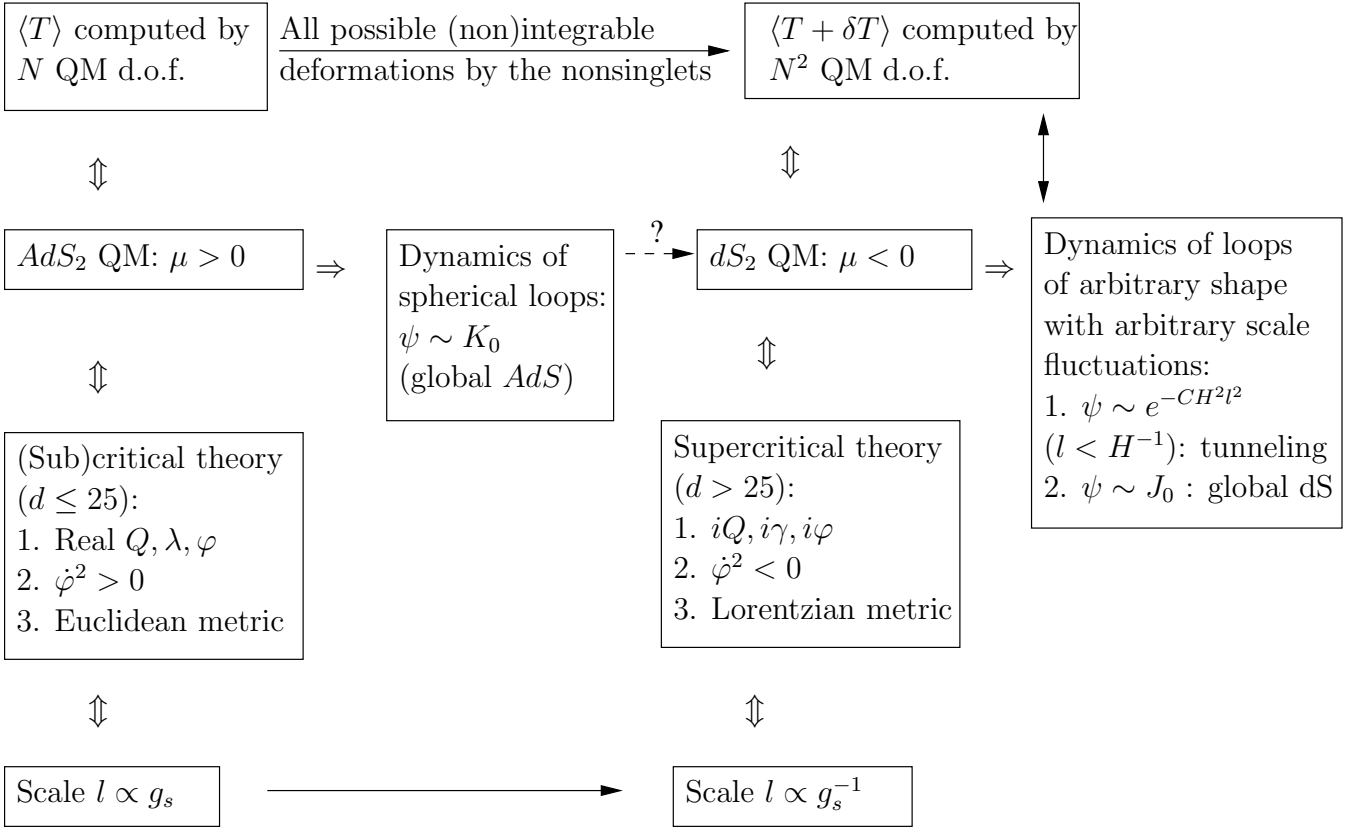} \caption{Summary of results and observations.}
\label{fchart-ch2}
\end{figure}


\section{The large $N$ RG in MQM}
\setcounter{equation}{0}

We will now review in some detail the steps of the large $N$ world
sheet RG analysis in $MQM$. For a complete description see the
original works in \cite{ts-s,ts-ns}. The concept arises from the
interpretation of the very existence of the {\it double scaling
limit} (or the continuum limit) as some kind of Wilsonian RG flow
\cite{bz}.

In the double scaling limit, as the matrix coupling constant $g \to
g_c$, the average number of triangles in triangulations at any genus
$G$ diverges as

\be \langle n_G \rangle \sim (1-G)(\gamma_0-2)(1-g/g_c)^{-1} \,.\ee
Simultaneously with $N \to \infty$, the length of the triangles (the
regularized spacing of the random lattice)  $a\sim
N^{-\frac{1}{2-\gamma_0}}$ scales to zero to keep the physical area
$a^2 \langle n_G \rangle \sim N^{-\frac{2}{2-\gamma_0}} (1-g/g_c)$
or equivalently the string coupling $g_s \equiv N^2
(g-g_c)^{2-\gamma_0}$ fixed. The existence of double scaling limit
indicates that a change in the length scale induces flow in the
coupling constants of the theory in a way that one reaches the
continuum limit with correct scaling laws and the critical exponents
at the nontrivial IR fixed point determined by the flow equations.
The large $N$ world sheet RG in $MQM$ \cite{ts-s, ts-ns} is
basically the evolution of the two sets of parameters of the theory,
the size of the matrices $N$ and the cosmological constant (mapped
into the matrix coupling $g$) and all other matrix couplings, at the
constant long distance physics with the rescaling of the
regularization length in the triangulation of the world-sheet. In
the Wilsonian sense this is done by changing $N\to N+\delta N$ by
integrating out some of the matrix elements, which is like
integrating over the momentum shell $\Lambda-d\Lambda < |p| <
\Lambda$, and compensating it by enlarging the space of the coupling
constants $g\to g+\delta g$. Here the space of coupling constants
will contain both the matrix coupling $g$ and the mass parameter
$M^2$.


\subsection{The RG scheme}

Let us now schematically explain the emergence of the flow
equations. By integrating out a column and a row of an $(N+1)\times
(N+1)$ matrix, the $c=1$ matrix partition function satisfies a
discrete relation

\be Z_{N+1}(g,M,R) = [\lambda(g,M,R)]^{N^2} Z_N(g',M',R')
\label{lambda-scaling} \ee with the following flow equations for the
couplings

\ba g' &=& g + \frac{1}{N}\beta_g(g,M,R)+O\Big(\frac{1}{N^2}\Big)\,,
\nonumber \\
M'^2 &=&
M^2+\frac{1}{N}\beta_{M^2}(g,M,R)+O\Big(\frac{1}{N^2}\Big)\,,
 \label{beta-defn} \ea and an auxiliary flow equation

\be \lambda(g,M,R)=1+N^{-1}~r(g,M,R)+O(N^{-2}) \,. \label{r-defn}
\ee Here $g', M'$ represent the renormalized couplings. Identifying
$1/N$ as the world sheet scale of the RG transformation, the
functions $\beta_g, \beta_M$ are interpreted to be the beta
functions for the corresponding couplings. The function $\lambda$
gives the change in the world sheet free energy (the string
partition function)

\be \mathcal{F}(N,g,M,R)=\frac{1}{N^2}\ln Z_N(g,M,R)
\label{string-partition} \,.\ee The auxiliary flow relation for
$\beta_{\lambda}$ is useful in studying thermodynamical
consequences. As $N$ is taken to be large, the world sheet scale
$1/N$ becomes infinitesimal. Then the discrete relation
(\ref{lambda-scaling}) for the Matrix partition function actually
translates into a continuous Callan-Symanzik equation satisfied by
the world sheet free energy with the function $r$ controlling it's
inhomogeneous part

\be \Big[N\frac{\partial}{\partial
N}-\beta(g,M,R)\frac{\partial}{\partial
g}-\beta(g,M,R)\frac{\partial}{\partial M}+\gamma \Big]
\mathcal{F}(N,g,M,R)=r(g,M,R) \,. \label{C-S} \ee Here $\gamma$ acts
as the anomalous dimension. We will see that the $\beta_g, \beta_M$
computed by the large $N$ RG are such that the homogenous part of
the Callan-Symanzik equation indeed determines the correct scaling
exponents for $c=1$ matrix model around the nontrivial fixed point.
The inhomogeneous part is related to some subtleties in the theory,
like the {\it logarithmic scaling violation} of the $c=1$ matrix
model.


\subsection{Integrating over the vectors (bosonic quarks)}

To carry out the explicit derivation of the flow equations, let us
consider the usual hermitian matrix quantum mechanics for
$(N+1)\times (N+1)$ matrices $\phi_{N+1}(t)$ with a periodic
boundary condition on the matrices. It has a cubic potential acting
as a wall stabilizing the inverted oscillator potential. Now
considering the following parametrization (the scalar being of
relative order $1/N$ is neglected in the double scaling limit),

\be \phi_{N+1}(t) = \begin{pmatrix} \phi_N(t) &  v_a(t) \cr v^*_a(t)
& 0 \cr \end{pmatrix} \,, \label{matrixpara}\ee and considering the
power expansion (the higher order terms in $v^*v$ are suppressed by
powers of $O(1/N)$),

\ba &&\mbox{Tr}\phi_{N+1}^{2k} =
\mbox{Tr}\phi_N^k+2k~v^*\phi_N^{2k-2} v+O(v^4) \,,
\nonumber\\
 &&\mbox{Tr}\phi_{N+1}^{2k+1} = \mbox{Tr}\phi_N^k+(2k+1)~
 v^*\phi_N^{2k-1} v+O(v^4\phi^{2k-3}) \,,
\ea the matrix partition function reduces to a part containing $N
\times N$ matrices and an integral over the vectors

\ba \mathcal{Z}_{N+1}[g,M,R] &=& \int_{\phi_N(2\pi R)=\phi_N(0)}
\mathcal{D}^{N^2}\phi_{N}(t) \nonumber \\ & &~\exp[-(N+1) \mbox{Tr}
\int_0^{2\pi R} dt~\{\frac{1}{2}{\dot \phi}_{N}^2(t) +
\frac{1}{2}M^2\phi_{N}^2(t)
-\frac{g}{3}\phi_{N}^3(t)\}] \nonumber \\
& & \int_{v,v^*(2\pi
R)=v,v^*(0)}\mathcal{D}^Nv(t)\mathcal{D}^Nv^*(t) \nonumber \\ &
&~\exp [-(N+1)\int_0^{2\pi R}dt~v^*(t)\{-\partial_t^2+M^2 - g\phi_N
(t)\}v(t)] \,.
\nonumber \\
& & \label{Z2_{N+1}} \ea This is nothing but the $c=1$ partition
function suitable for Veneziano type QCD considered in
\cite{yang-open,minahan-open} to study open strings. Both color $N$
and fermion (quark) flavor $N_f$ are taken to be large. Here the
quarks are bosonic (the vectors) and $N=N_f$. It is precisely these
fields in the fundamental representation of the global $SU(N)$
group, which generate boundary terms in the Feynman diagrams.
Integrating adiabatically over these (bosonic) quark loops generates
a logarithmic interaction term ($+N~\mbox{Tr}\log [-\partial_t^2+M^2
- g\phi_N (t)]$) and hence the determinant $\mbox{det}
||-\partial_t^2+M^2-g\phi_N||^{-1}$ that gives rise to the
boundaries in the world-sheet. Logarithm with minus sign would arise
if the integration is performed on $N$ flavors of fermions. Unlike
the open string models mentioned above, here the couplings in front
of the logarithm and in its argument are nor introduced by hand,
rather they are determined by the original closed string action. One
can ignore the derivative term inside the logarithm if the mass and
the couplings are large enough. This makes the dynamical loops
uncorrelated in time and gives rise to Dirichlet boundaries on the
world sheet. On the other hand, if the derivative term dominates
then the world sheet boundaries are Neumann. In \cite{minahan-open},
$c=1$ model with explicit expression for the fundamental fields has
been considered and is the one which is closest to our model
obtained after integrating out one row and one column of flavor
degrees of freedom. If one considers infinite line, only ground
state is relevant. For some choices of the coupling and the mass of
the particle moving around the boundary of the holes, it has been
possible to find the ground state \cite{minahan-open}, but the exact
spectrum is not known. Our ultimate goal is to derive the exact
spectrum of these boundaries and to study their cosmological
implications.


\subsection{Evaluating
$\mbox{det}||-\partial_t^2+M^2-g\phi_N||^{-1}$}

\begin{figure}[htb]
\epsfysize=5cm \centerline{\epsffile{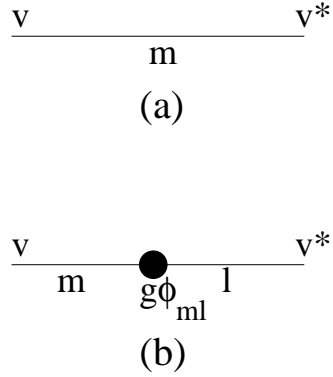}} \caption{The
propagators and vertices: (a) $[\mc{O}_{mn}^{v^*v}]^{-1}$, (b)
$[\mc{O}_{m-l}^{v^*v}(g,\phi)]^{-1}$.} \label{props}
\end{figure}

Let us now evaluate the determinant by standard Feynman expansion.
In general it is a nonlocal object. For simplicity one could
consider the constant $\phi_N$ mode, which is equivalent to studying
the effective potential to determine the phase structure. But here
we will treat more general case of evaluating the determinant with
flavors coupling to general $\phi_N$. The complicated induced
interactions arising from the logarithmic term are ignored in small
field approximation in order to get back the beta functions that
depend only on the original couplings. In terms of the Fourier
modes, the $v$-integration can be expressed as

\be I[g,M,\phi_N,R]=\frac{1}{[\sqrt{2\pi R(N+1)}]^{2N}}\int
\big(\prod_n dv_n^*dv_n^*\big) ~\exp[-\sum_m
v_m^*\big(\frac{m^2}{R^2}+M^2\big)v_m
+g\sum_{m,l}v_m^*\phi_{m-l}v_l] \,. \ee The integral can be
re-casted in terms of a gaussian part $I_0$ being acted upon by a
part generating interaction vertices

\be I[g,M,\phi_N,R,N] = \frac{1}{[2\pi
R(N+1)]^{N}}~\exp\Big[-\sum_{m_1,l_1}
\mc{O}_{m_1-l_1}^{v^*v}(g,\phi)\Big]~I_0(R,M,N)\,, \label{I1} \ee
with the gaussian part defined as

\be I_0[R,M,N]=\int \Big(\prod_jdv_j^*dv_j\Big) ~\exp\big[-\sum_m
v_m^*\mc{O}_{mm}^{v^*v}v_m\big]\,. \ee Here the operators
$\mc{O}_{mn}^{v^*v}$ and $\mc{O}_{m-l}^{v^*v}(g,\phi)$ respectively
define the inverse of propagator carrying a momentum $mn/R
\delta_{mn}$ and interaction vertex carrying a net momentum
$(m-l)/R$

\ba \mc{O}_{mn}^{v^*v} &=&
\Big(\frac{mn}{R^2}+M^2\Big)\delta_{mn}\,,~~~
 \mc{O}_{m-l}^{v^*v}(g,\phi) = g\phi_{m-l} \,.
\label{propagators} \ea The inverse of these operators define
various propagators and vertices according to figure \ref{props}.

The gaussian integration is basically absorbed in a net prefactor in
(\ref{I1}) such that it reduces to the sum of one loop Feynman
diagrams $\Sigma[g,M,\phi_N,R,N]$ up to this prefactor.
The sum $\Sigma [g,M,\phi_N,R,N]$ is given by

\begin{figure}[htb]
\epsfysize=3cm \centerline{\epsffile{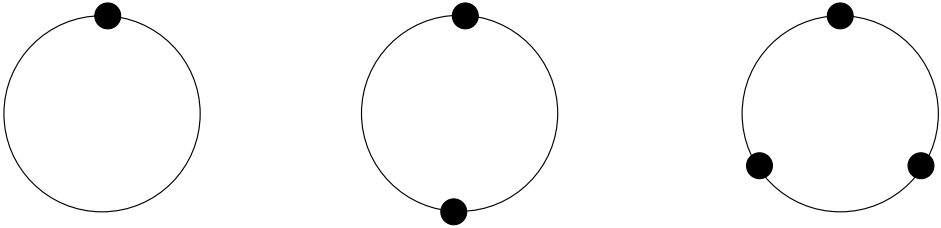}} \caption{The
diagrams contributing to $\Sigma$ at one-loop order.} \label{diags}
\end{figure}

\ba &&\Sigma[g,\phi_N,R,N] = 1 + g~\mbox{Tr}\bigg[\sum_n\frac{1}
{\frac{n^2}{R^2}+M^2}~\phi_0\bigg]
\nonumber \\
&&+\frac{g^2}{2}~\mbox{Tr}\bigg[\sum_{m,n}\frac{1}{\big(\frac{m^2}{R^2}+M^2
\big)\big(\frac{n^2}{R^2}+M^2\big)}~\phi_{m-n}\phi_{n-m}\bigg]
\nonumber \\
&&+\frac{g^3}{3!}~\mbox{Tr}\Bigg[\sum_{m,n,l}\frac{1}
{\big(\frac{m^2}{R^2}+M^2\big)\big(\frac{n^2}{R^2}+M^2\big)
\big(\frac{l^2}{R^2}+M^2\big)}~\phi_{m-l}\phi_{l-n}\phi_{n-m}
\Bigg]+O(g^4) \label{SigmaFourier} \ea Now the one loop correction
to the various coefficients in the effective action is evaluated
term by term by inverse transforming the Fourier modes, summing up
the set of infinite series and then exponentiating and log-expanding
$\Sigma[g,M,\phi_N,R,N]$ using small field approximation. The
summation is done according to

\ba &&\sum_{m=-\infty}^{\infty}
\frac{\exp[i(m/R)t]}{\frac{m^2}{R^2}+M^2} = \frac{\pi
R}{M}\frac{\cosh(\pi MR-Mt)}{\sinh\pi MR} \,,~~~0\le t\le 2\pi R
\,.\ea Since after performing the inverse Fourier transform, the
various terms has nonlocal integrals over several one dimensional
dummy time variables, we breakup the variables into center of mass
and relative coordinates. Then we expand the functions about the
center of mass coordinates, assuming the relative coordinates to be
small enough, and consider integration over the relative
coordinates. The higher derivative terms are dropped due to
smallness of the relative coordinate compared to the center of mass
coordinate. For example, by the coordinate transformation $(t_1,
t_2) \to (T=(t_1+t_2)/2, \tau=(t_1-t_2)/2)$, the $O(g^2)$ term of
the form $\int dt_1 \int dt_2 ~\mbox{Tr} [\phi(t_1) \phi(t_2)]$ can
be expanded around $T$, considering $\tau$ small (which is true for
$t_2 >0$). Then the $\dot \phi^2$ term comes from $O(\tau^2)$ term
in the expansion, whereas the next higher derivative term appears at
$O(\tau^4)$. All the mixed terms of powers of $\phi$ and the powers
of its derivatives appear in odd orders of $\tau$ and hence are
automatically dropped due to integration of $\tau$ over a symmetric
range.

Collecting all the contributions up to $O(\phi\phi\phi)$, the
expression for $\Sigma[g,M,\phi_N,R,N]$ becomes

\ba \Sigma[g,M,\phi_N,R,N] &\simeq& 1+ F_{g1}(R,M)~g \int_0^{2\pi R}
dt~\mbox{Tr}\phi_N(t) + F_{g2}(R,M)~g^2 \int_0^{2\pi R}dt
~\mbox{Tr}\phi_N^2 (t)
\nonumber \\
&&+\hat{F}_{g2}(R,M)~g^2\int_0^{2\pi R}dt
~\frac{1}{2}\mbox{Tr}\dot\phi_N^2(t) +F_{g3}(R,M)~g^3\int_0^{2\pi
R}dt~\frac{1}{3}\mbox{Tr}\phi_N^3(t)\,. \nonumber\\
\label{SigmaInvFourier} \ea Note that, in the evaluation of $\Sigma$
we keep the contribution from the nonlocal terms of the action up to
the kinetic term and the contribution from the higher order terms in
the matrix field up to the cubic term. All other operators are
redundant for our purpose and are negligible due to the small field
approximation. The coefficients $F(R)$s are defined as follows

\ba F_{g1}(R,M)&&= \frac{1}{2M} \coth \pi MR\,,
\nonumber \\
F_{g2}(R,M)&&= \frac{1}
 {M^3 \sinh ^2 \pi MR }\Big( \frac{1}{2}\pi M R \cosh 2\pi MR+
 \frac{1}{8}\sinh 4 \pi
 MR \Big)\,,
 \nonumber\\
\hat F_{g2}(R,M) &&= \frac{1}{M^5 \sinh^2 \pi MR}\Big(\frac{\pi M
R}{16}\cosh 4\pi M R-\frac{\pi^3 M^3 R^3}{6} \cosh 2\pi MR
\nonumber\\
&&-\frac{1}{64}(1+8 \pi^2 M^2 R^2)\sinh 4\pi M R \Big) \,,
\nonumber\\
F_{g3}(R,M) &&=\frac{\pi M R}{64 M^5 \sinh^3\pi M R (\cosh 2 \pi M R
+ \cosh 4 \pi M R ) }
\nonumber \\
&&[ 4 \pi M R \big(3 \cosh \pi M R + 2 \cosh3 \pi M R +2 \cosh 5 \pi
M R + \cosh 7 \pi M R \big)
\nonumber \\
&&+ \sinh \pi M R + \sinh 3 \pi M R+ \sinh 5 \pi M R+ 2 \sinh 7 \pi
M R + \sinh 9 \pi M R ]\,.
\nonumber \\
\label{defhyperbolic} \ea The tadpole term linear in $\phi_N(t)$ is
now removed by shifting the background and setting it's net
coefficient to zero. This fixes the value of the shift ( $f\sim-g
M^2 F_{g1}/N + O(\frac{1}{N^2})$) and accordingly modifies the
coefficients of all the terms in the action.


\subsection{Restoring the original cut-off and
the space-time geometry around a fixed point}

As usual in Wilsonian RG we now restore the original cut-off by
rescaling the variables

\be \phi_N(t) \to \rho\phi'_N(t')\,,~~~t' \to t(1-h~dL)\,,~~~R' \to
R(1-h~dL)\,.\label{rescaling}\ee Here $dL=1/N$ and $h$ is a function
of the radius and the couplings whose functional form can be
determined from the behavior of the flow equations near the fixed
points. As one approaches a fixed point, $h$ saturates to a constant
characteristic to that particular fixed point, in the sense that it
determines the scaling exponents. Thus the rescaling of the radius
$R$ in some sense tells us that as the system flows to various fixed
points, the target space geometry changes accordingly. This is
something new in matrix model which directly enables us to determine
the target space metric around a fixed point by solving the equation
in it's neighborhood \be \frac{d R}{dL}= -h R\,.\label{betaR}\ee For
example, later in this section we will see that $h=0$ corresponds to
a $c=1$ fixed point. This trivial rescaling indeed gives a flat
metric ($R^2=const$). In \cite{ts-bh}, this principle is used to
extract $2D$ Euclidean black hole metric from matrix quantum
mechanics.


\subsection{Flow equations and the nontrivial fixed point}

We now proceed towards determining the flow equations by setting the
overall coefficient of the kinetic term to one. This gives $\rho$ as
\be \rho =1+\frac{1}{2}(h-1+g^2 \hat F_{g2}) dL + O(dL^2)\,.\ee The
resulting effective action is of the same form as the bare one with
the renormalized strength of the couplings. The renormalized
partition function is given by

\ba Z_{N+1} &=& \lambda'~^{N^2}
\int\mc{D}^{N^2}\phi'_N(t')~\exp\Big[-N~\mbox{Tr}\int_0^{2\pi
R'}dt'\Big(\frac{\dot{\phi'_N}^2(t')}{2}+M'^2\frac{{\phi'_N}^2(t')}{2}
-g'\frac{{\phi'_N}^3(t')}{3}\Big)\Big] \,,\nonumber \\
g' &=& g+\Big(\frac{5}{2}h-\frac{1}{2}\Big)g~dl+\big[F_{g3} (R,M)
-\frac{3}{2}\hat{F}_{g2}(R,M)\big]g^3~dL\,, \nonumber \\
M'^2 &=&
M^2+\big[2hM^2+g^2(1-M^2)F_{g2}(R,M)-g^2M^2\hat{F}_{g2}(R,M)\big]dL\,.
\ea The resulting beta function equations are

\ba \beta_g &=& \frac{d g}{dL} = \Big(\frac{5}{2}h-\frac{1}{2}\Big)g
+ \big[F_{g3}(R,M)
-\frac{3}{2}\hat{F}_{g2}(R,M)\big]g^3 \,, \nonumber \\
\beta_{M^2} &=& \frac{dM^2}{dL} = 2hM^2 +g^2[(1-
M^2)F_{g2}(R,M)-M^2\hat{F}_{g2}(R,M)]\,.\label{bfe}\ea As we already
mentioned much of the structure of the beta functions depend on
understanding the quantity $h$. The gaussian fixed point
$\Lambda^{*}_{1}=\big(0,0\big)$ satisfies both the equations
$\beta_g=0$ and $\beta_{M^2}=0$ trivially for any $h$. One can show
that for $h=0$, or in other words for trivial rescaling ($t'=t$,
$R'=R$) of the target space coordinates, the nontrivial fixed points
$\Lambda^{*}_{2}(g^{*2} \ne 0, M^* \ne 0)$ have $c=1$ critical
exponents. Such a pair of nontrivial $c=1$ fixed points of the flow
equations (\ref{bfe}) exists for any $R$ and are given by

\ba &&g^{*2}={\frac{1}{2F_{g3}(R,M^*)-3\hat
F_{g2}(R,M^*)}}\,,\nonumber\\&&(1-
M^{*2})F_{g2}(R,M^*)-M^{*2}\hat{F}_{g2}(R,M^*)=0 \label{fp}\,.\ea
The second equation solves for the values of $M^*$ for different
$R$. The fixed point values of the couplings are scheme dependent.
As such their explicit values do not affect the physics as long as
they are finite. However, to study the Dirichlet boundaries on the
world sheet, we are interested in large values of $M^*$. As one cuts
a hole on the triangulated world sheet, there is no longer any
matter or the matrix valued fields sitting inside the cut portion.
Instead there are vectors $v, v^*$ who sew the cut portion by
correlating the matter at the different points on the boundary. As
their propagator goes like $1/M^*$, a large $M^*$ would imply
uncorrelated matter on the boundary and hence will serve as a
Dirichlet boundary condition \cite{minahan-open}. For $h=0$, solving
the corresponding equation for $M^*$ (\ref{fp}), we find that a
large $M^*$ is feasible for $\pi R \le 1$. This implies that the
effects of nonsinglet sector cannot be suppressed in this situation.
In fact $M^*$ can have any value at vanishing $R$ and is of order
one in the range $0< \pi R <1$. Around $\pi R \sim 1$ it goes to a
peak ($M^* \sim 2$) and then rapidly goes to zero.


\subsection{The $c=1$ critical exponent}

Let us now explain how a nontrivial fixed point of (\ref{bfe}) with
$h=0$ corresponds to a $c=1$ fixed point. The critical exponents for
the scaling variables, the renormalized bulk cosmological constant
$\Delta=1-g/g^*$ and the renormalized mass (related to the
renormalized boundary cosmological constant) $\mc{M} = 1-M/M^*$ can
be determined from the eigenvalues of the matrix

\be \Omega_{k,l} = \frac {\partial \beta_{k}(\Lambda^*)} {\partial
\Lambda_{l} }\,. \ee The homogeneous part of the Callan-Symanzik
equation, satisfied by the singular part of the free energy, can now
be rewritten as

\be \Big[N \frac{\partial}{\partial N}-\Omega_{1} ~\Delta
\frac{\partial}{\partial \Delta} -\Omega_{2}~\mc{M}
\frac{\partial}{\partial~\mc{M}}+2 \Big]
\mc{F}_{s}~[~\Delta,\mc{M},R] = 0 \label{cs-scaling}\,. \ee The
eigenvalues $\Omega_{i}$s  of the matrix $\Omega_{k,l}$ are nothing
but the scaling dimensions of the relevant operators. The singular
behavior with respect to the renormalized bulk cosmological constant
goes as, \be \mc{F} _{s} \sim \Delta ^{2/\Omega_{1}} f_{1} [
N^{\Omega_{1}}~\Delta~]~f_{2} [ N^{\Omega_{2}} ~\mc{M}~]\,. \ee
Comparing the above expression of $\mc{F}_{s}$ with the matrix model
result ~$\mc{F}_{s} \sim \Delta ^{(2-\gamma_{0})} ~f[
N^{2/\gamma_{1}} \Delta]$, or using the standard definition of the
susceptibility ~$\Gamma ~\sim \frac {\partial^2 \mc{F}_{s}}
{\partial \Delta^2} \arrowvert_{~\mc{M}=0} ~\sim \Delta
^{-\gamma_{0}}~$, the string susceptibility exponent $\gamma_{0}$ is
given by

\be \gamma_{0} \sim (2-2/\Omega_{1})\,. \ee Note that in our
analysis  $\gamma_1 \sim 2/\Omega_{1}$ is consistent with the matrix
model relation $\gamma_0 + \gamma_1 = 2$. This relation is
independent of the explicit values of $\gamma_0$ and $\gamma_1$ and
is easily obtainable from the consideration of the torus. The string
susceptibility exponent at genus $G$ is defined by

\be \gamma_G = \gamma_0 + G~\gamma_1 \,.\ee  For the nontrivial
fixed point, $\Omega_{11}=1+5h/2$. Note that in this RG, typically
$g^*$ is small. The pair of nontrivial fixed points is always
situated close to the Gaussian fixed point. This implies the
off-diagonal elements $\Omega_{12},~\Omega_{21}~$, being proportional
to the powers of $g^*$, are small and $\Omega_{22} \sim h$. The
scaling dimension matrix is effectively a diagonal one with
$\Omega_1=1+5h/2,~\Omega_2=h$. Thus $h=0$ gives the $c=1$ critical
exponent, $\gamma_{0}=0$. This is true for any $R$. The pair of
$c=1$ fixed points we thus get are repulsive with respect to the
flow of the parameter $g$ while the gaussian fixed point is
attractive. One can show that this pair of fixed points reproduces
all the known physics of the $c=1$ theory and respects $T$-duality.


\section{The AdS minisuperspace from large $N$ RG}
\setcounter{equation}{0}

We would like to know that in the context of our large $N$ RG, which
is formally dealing with the full superspace, is there really any
limit in terms of the scale $l$ that would confine us to the
minisuperspace and reproduce the familiar minisuperspace results
computed by the continuum Liouville theory or by the free fermionic
matrix model? We would also like to know that if such a limit
exists, then how the full quantum wave function is differing from
the minisuperspace wave function away from this limit. This will
help us in understanding the limitations of such a truncation of
superspace in computing the actual quantum wave function of the
universe. We will see that the minisuperspace physics in $c=1$
matrix model corresponds to quantum mechanics of $N$ degrees of
freedom, the eigenvalues. Here, we will show that the large $N$ RG
acting on matrix quantum mechanics observables on the space of
eigenvalues correctly captures the familiar minisuperspace WdW
equation. This tests the correctness of the approach in deriving the
WdW constraint.

To be more specific, let us consider the one point function of loop
operator in the space of $N+1$ eigenvalues from the diagonal
matrices $\Lambda_{N+1}(\hat x) = \Omega^{-1}\phi_{N+1}(\hat
x)\Omega ,\, \Omega \in SU(N)$ with periodic boundary condition
$\lambda(0)=\lambda(2 \pi R)$, given by \be \langle\mbox{Tr}e^{l
\lambda_{N+1}(\hat x)}\rangle=\langle e^{l \lambda(\hat x)}
\rangle^{N+1}\,.\ee We then integrate out one eigenvalue
$\lambda_0(\hat x)$ adjusting the matrix couplings according to
double scaling limit, thus keeping $g_s$ fixed. This generates a
difference equation

\be \langle\mbox{Tr}e^{l \Lambda_{N+1}(\hat
x)}\rangle_{\eta_i}-\langle\mbox{Tr} e^{l\Lambda_N(\hat
x)}\rangle_{\eta_i'}=(\langle e^{l \lambda_0(\hat
x)}\rangle_{\eta_i'}-1)\langle\mbox{Tr}e^{l \Lambda_N(\hat
x)}\rangle\,, \label{ads-qm}\ee where, $\eta_i'$ are the
renormalized matrix
couplings given by \ba &&g'=g(1-1/2N)\,,~~\beta_g=-g/2\,,\nonumber\\
&&M'^2=M^2\,,~~\beta_{M^2}=0\,. \ea Note that, the beta functions
above are too trivial to capture the nontrivial $c=1$ fixed point
and thus only have the trivial gaussian fixed point. This happens
because in the space of the eigenvalues one practically works with
the gaussian matrix model.

Now the left hand side of the difference equation can be interpreted
as the Callan-Symanzik operator

\be \langle\mbox{Tr}e^{l \Lambda_{N+1}(\hat
x)}\rangle_{\eta_i}-\langle\mbox{Tr} e^{l \Lambda_N(\hat
x)}\rangle_{\eta_i'}=\Big(\frac{\partial}{\partial N}-\delta g
\frac{\partial}{\partial g}\big) \langle\mbox{Tr} e^{l
\Lambda_N(\hat x)}\rangle_{\eta_i'}\,,\ee that simplifies to

\ba &&\Big(\frac{\partial}{\partial N}-\delta g
\frac{\partial}{\partial g}\Big) \langle e^{l \lambda(\hat
x)}\rangle_{\eta_i'}^N\nonumber\\&&=N \Big\langle-\int dt
\big[\lambda (t)\frac{1}{2}(-\partial_t^2+M^2)\lambda(t)-g
\lambda^3(t)\big]e^{l \lambda(\hat x)}\Big\rangle_{\eta_i'}~\langle
e^{l \lambda(\hat x)}\rangle_{\eta_i'}^{N-1} \,.\ea Now, just like
in free fermionic theory, rescaling $\lambda_N \to
\frac{\lambda_N}{\sqrt{N}}$ cubic interaction term becomes
$O(1/N^2)$ and we focus on the top of the quadratic potential.
Simultaneously we rescale $l \to \sqrt{N} l$ keeping the arguments
of the exponentials unchanged. This makes

\be \Big(\frac{\partial}{\partial N}-\delta g
\frac{\partial}{\partial g}\big) \langle\mbox{Tr} e^{l
\Lambda_N(\hat x)}\rangle_{\eta_i'}=-\frac{1}{2}M^2 \partial_l^2
\langle\mbox{Tr} e^{l \Lambda_N(\hat x)}\rangle_{\eta_i'}\,.\ee
Plugging this in (\ref{ads-qm}) we have the minisuperspace WdW
equation

\be [-\frac{1}{2} M^2 l^2 \partial_l^2+4\mu l^2]
\psi(l)=0\,.\label{ads-qm-f} \ee Here the wave function $\psi(l)$ is
given by the one point function $\langle\mbox{Tr} e^{l
\lambda_N(\hat x)}\rangle_{\eta_i'}$. Comparing (\ref{ads-qm-f})
with the standard minisuperspace WDW equation (\ref{minisupwdw}),
the world sheet cosmological constant is given by

\be 4 \tilde \mu= 8 \mu /M^2=\frac{2}{M^2}(1-\langle e^{l
\lambda_0}\rangle_{\eta_i'})\,. \ee Now to determine the sign of the
cosmological constant, recall \cite{gm, kleb, mart} that in the free
fermionic theory the single particle wave functions in the inverted
harmonic oscillator potential are supported at $- \infty \le \lambda
\le 0$ (note that, here we have opposite sign of the cubic
interaction term as that of \cite{mart}). Also the cosmological
constant being of $O(1/M^2)$ is small. Thus $\tilde \mu \gtrsim 0$
($\tilde \mu \to 0$ as $\lambda_0 \to 0$ and $g \to 0$) and
comparing (\ref{ads-qm-f}) with (\ref{minisupwdw}), we see that the
large $N$ RG of the matrix model observables defined on the space of
eigenvalues indeed gives rise to $AdS$ quantum mechanics, as would
be expected from free fermionic theory. This tests the validity of
our approach for the WdW cosmology we are addressing in this paper.


\section{The modified Wheeler-de Witt constraint}
\setcounter{equation}{0}

So far we have discussed the setup for large $N$ RG analysis giving
rise to flow equations that have nontrivial $c=1$ fixed points. We
also have discussed the minisuperspace limit of the large $N$ RG
flow that gives rise to the familiar $AdS$ QM computed by free
fermion representation of the $c=1$ matrix model. Let us now move
towards the computation of the wave function by deriving the
modified Wheeler-de Witt  (WdW) equation that dynamically determines
the required wave function. In fact, the Callan-Symanzik equation
for the one point function of the loop operator with Dirichlet
boundary condition itself gives rise to this constraint.


\subsection{Integrating over the vectors}
Let us consider the expectation value of the loop operator at a
fixed $\hat x$ with respect to the hermitian matrix quantum
mechanics path integral with a cubic potential and couplings $\eta_i
= g, M^2$ \be \psi(l)= \frac{1}{N}\big \langle \mbox{Tr}
~e^{l\phi(\hat x)}\big \rangle_{\eta_i} \,.\label{wavefn}\ee As
before, the matrices obey periodic boundary condition. Considering
the parametrization (\ref{matrixpara}) and expanding the operator,
we have

\be \langle \mbox{Tr} ~e^{l\phi_{N+1}(\hat x)}\rangle_{\eta_i} =\int
D\phi_N (t) ~e^{-\tilde S_N(\phi)}I_{v(t)}\sum_{k=0}^\infty
\frac{l^k}{k!} [\mbox{Tr}\phi_N^k(\hat x)+kv_N^*(\hat
x)\phi_N^{k-2}(\hat x)v_N(\hat x)]\,.\ee Here $\tilde S_N (\phi)$ is
the $v,v^*$ independent part of the reduced action given by

\be \tilde S_N (\phi)= N~\mbox{Tr}
(1+\frac{1}{N})\Big[\frac{\dot\phi_N^2}{2}
+\frac{M^2}{2}\phi_N^2-\frac{g}{3}\phi_N^3\Big]\,.\ee The integral
over the vectors $I_{v(t)}$ is the usual one as in (\ref{Z2_{N+1}})
\be I_{v(t)} = \int dv_N^*(t)dv_N(t) \exp[-\int_0^{2\pi
R}dt~v_N^*(t)(-\partial_t^2+M^2-g\phi_N(t))v_N(t)] \,. \label{Iv}\ee
As we have discussed already, this is basically similar to the
integral over the bosonic quark loops in Veneziano type QCD with a
large color and flavor ($N=N_f$). The change in the expectation
value of the loop operator due to the renormalization of the
couplings $\eta_i$ can now be expressed as

\be \langle\mbox{Tr}e^{l\phi_{N+1}(\hat x)}\rangle_{\eta_i}
-\langle\mbox{Tr}e^{l\phi_N(\hat x)}\rangle_{\eta_i'}=
l\Big\langle\frac{1}{\Sigma} ~I_{v(t)}v^*_N(\hat x)e^{l\phi_N(\hat
x)}\phi_N^{-1}(\hat x)v_N(\hat x)\Big\rangle_{\eta_i'}\,.
\label{exptn-diff}\ee Here $\Sigma$ represents the diagrammatic
expansion of $I_{v(t)}$ that eventually renormalizes the bare action
(recall  (\ref{SigmaInvFourier})). Thus the expectation value of any
operator $\mathcal{O}$ with respect to the renormalized couplings
$\eta_i'$ can formally be given by

\be \langle\mathcal{O}\rangle_{\eta_i'} = \int D\phi_N~\mathcal{O}
~e^{-\tilde S_N [\phi]}~\Sigma \,.\ee We will now closely examine
the meaning of the both sides of the equation (\ref{exptn-diff}).


\subsection{The Callan-Symanzik operator and the modified WdW constraint}

The change in the expectation value in (\ref{exptn-diff}) is
essentially the Callan Symanzik operator in discrete form, acting on
the one point function of the loop operator. Consider the discrete
relation

\be \langle\mbox{Tr}e^{l\phi_{N+1}(\hat x)}\rangle_{\eta_i}
-\langle\mbox{Tr}e^{l\phi_N(\hat
x)}\rangle_{\eta_i'}=\Big[\frac{\partial}{\partial N}-\delta \eta_i
\frac{\partial}{\partial \eta_i} \Big]
\langle\mbox{Tr}e^{l\phi_{N}(\hat x)}\rangle_{\eta_i}\,.\ee Now
recalling (\ref{beta-defn}) defining the beta functions
$\beta_{\eta_i}=N \delta \eta_i$ in the large $N$ RG, and using the
identity $\frac{\partial}{\partial N} \langle \mbox{Tr}
~e^{l\phi_{N}(\hat x)} \rangle_{\eta_i}=(N
\partial/\partial N+1 ) \frac{1}{N} \langle \mbox{Tr}
~e^{l\phi_{N}(\hat x)} \rangle_{\eta_i}$, the left hand side of
(\ref{exptn-diff}) can clearly be written as a Callan-Symanzik
operator acting on $\psi(l,\hat x)$

\be \langle\mbox{Tr}e^{l\phi_{N+1}(\hat x)}\rangle_{\eta_i}
-\langle\mbox{Tr}e^{l\phi_N(\hat x)}\rangle_{\eta_i'}=\Big[N
\frac{\partial}{\partial N}-\beta_{\eta_i} \frac{\partial}{\partial
\eta_i}+1 \Big] \frac{1}{N}\langle\mbox{Tr}e^{l\phi_{N}(\hat
x)}\rangle_{\eta_i}\,.\label{cso}\ee However, the wave function is
not completely annihilated by the Callan-Symanzik operator. The
nonvanishing right hand side is essentially computing the anomaly in
a nontrivial way. The origin of this anomaly lies in the fact that
the loops of same length can continuously be mapped into each other
by $2D$ diffeomorphism which acts like a gauge symmetry. The WdW
operator acts as a gauge fixing condition.

We will compute the anomaly below and will see that it is
proportional to $\phi_N^{-1}(\hat x)$ which is singular in the small
field approximation. The singularity can be tackled by taking
derivatives with respect to $l$ on both sides of the Callan-Symanzik
equation. The right hand side then simplifies to a differential
operator acting on the one point function. In this form, the
operator, which annihilates the one point function of the loop
operator, gives nothing but the {\it modified WdW constraint}.
Physically this constraint governs the dynamics of loops of
arbitrary shape with arbitrary length fluctuations as compared to
the usual (minisuperspace) WdW constraint that only deals with the
dynamics of a spherical loop. We will discuss in detail in what
sense this constraint is `modified', {\it i.e.} `beyond
minisuperspace', compared to the usual minisuperspace WdW constraint
and it's implications. In this paper, we will show that the very
existence of such a general `beyond minisuperspace' constraint in
low dimensional noncritical string theory is the key to probe
question like inflation and topology change.


\subsection{Computing the anomaly}

Let us now go back to the computation of the anomaly term. As
before, let us rescale the vectors as $v_i \to v_i/\sqrt{2\pi R}$
and Fourier decompose all the vector and matrix degrees of freedom.
We now perform the following small field expansion of the
non-gaussian part of (\ref{Iv}) containing the interaction vertex
$\mathcal{O}_{m-l}^{v^*v}(g,\phi)=g\phi_{m-l}$

\ba &&l\Big\langle\frac{1}{\Sigma}~I_{v(t)}v^*(\hat
x)e^{l\phi_N(\hat x)}\phi_N(\hat x)v(\hat x)\Big\rangle_{\eta_i'} =
l\Big\langle\frac{1}{\Sigma}~\int\frac{\prod_mdv_m^*dv_m}{(2\pi
R)^{N+1}}~\sum_{k,l}e^{\frac{i(-k+l)}{R}\hat x}v_k^*(e^{l\phi_N(\hat
x)}\phi_N^{-1}(\hat x))v_l \nonumber \\
&&\Big(1+g\sum_{m,n}v_m^*\phi_{m-n}v_n+\frac{g^2}{2!}\sum_{m,n}
\sum_{p,q}v_m^*\phi_{m-n}v_n~v_p^*\phi_{p-q}v_q+\ldots\Big)~\exp[-\sum_m
v_m^*(m^2/R^2+M^2)v_m] \Big\rangle_{\eta_i'} \,.\nonumber \\ \ea We
then contract all possible $v, v^*$ pairs in the expansion that
(upon acting on the remaining gaussian part of the integration)
extracts inverse propagator $(\mathcal{O}^{v^*v}_{mn})^{-1}=
(\frac{mn}{R^2}+M^2)^{-1}\delta_{m,n}$. The gaussian integration
$I_0$ gives an overall pre-factor that neutralizes a part of the
denominator. The contractions nontrivially mixes up the momentum
modes at interaction vertices generating complicated interaction
terms in the diagrammatic expansion. However, the Fourier modes of
the singular term $e^{l\phi_N(\hat x)}\phi_N^{-1}(\hat x)$ do not
get mixed with that of other terms. Hence the term is effectively
decoupled from the rest of the expansion and does not receive any
correction. The rest of the term with quantum corrections can
finally be cast as a series summation of differential operators
which we will explain below. This decoupling property of the
singular term eventually gives back the one point function of the
loop operator and is thus crucial in closing the WdW constraint.
Here for simplicity we do not show the Fourier decomposition of the
term. The resulting diagrammatic expansion due to the integration
over vectors with insertion of the loop operator at the Dirichlet
boundary is

\begin{figure} [ht]
\centering
\input{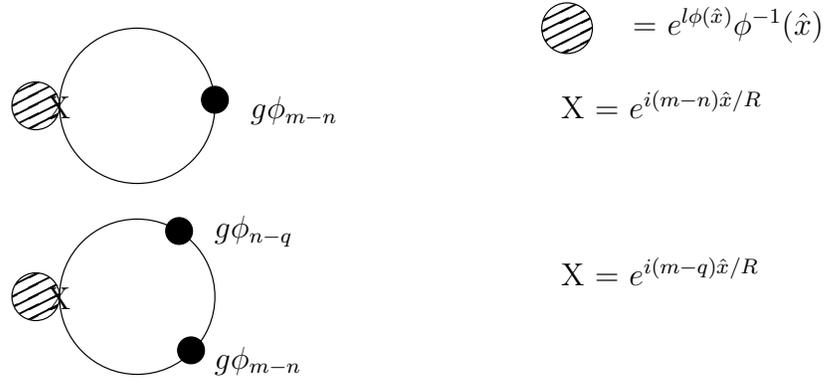} \caption{Diagrams
representing $O(g)$ and $O(g^2)$ terms
in (\ref{anomaly-expn1}).}
\label{feynman1}
\end{figure}

\ba &&\langle\mbox{Tr}e^{l\phi_{N+1}(\hat x)}\rangle_{\eta_i}
-\langle\mbox{Tr}e^{l\phi_N(\hat x)}\rangle_{\eta_i'} =\frac{l}{2\pi
R}\Big\langle\frac{1}{\tilde\Sigma}\Big[\mbox{Tr}\Big(\frac{\pi
R}{M}~\coth\pi MR~e^{l\phi_N(\tilde x)}\phi_N^{-1}(\hat x)\Big)
\nonumber \\ &&+g~\mbox{Tr}\Big(e^{l\phi_N(\hat x)}\phi_N^{-1}(\hat
x)\sum_{m,n}\frac{e^{i(-n+m)\hat
x/R}\phi_{m-n}}{(n^2/R^2+M^2)(m^2/R^2+M^2)}\Big) \nonumber \\
&&+\frac{g^2}{2}~\mbox{Tr}\Big(e^{l\phi_N(\hat x)}\phi_N^{-1}(\hat
x)\sum_{m,n,q}\frac{e^{i(-q+m)\hat
x/R}\phi_{m-n}\phi_{n-q}}{(m^2/R^2+M^2)(n^2/R^2+M^2)(q^2/R^2+M^2)}\Big)
\nonumber \\ &&+\frac{g^3}{6}~\mbox{Tr}\Big(e^{l\phi_N(\hat
x)}\phi_N^{-1}(\hat x)\sum_{m,n,q,k}\frac{e^{i(-k+m)\hat
x/R}\phi_{m-n}\phi_{n-q}\phi_{q-k}}
{(m^2/R^2+M^2)(n^2/R^2+M^2)(q^2/R^2+M^2)(k^2/R^2+M^2)} \Big)
\nonumber \\ &&+\ldots\Big]\Big\rangle_{\eta_i'}\,, \nonumber\\
\label{anomaly-expn1}\ea where $\tilde\Sigma$ denotes the
non-Gaussian part of the diagrammatic expansion of the vector
integral

\be \Sigma = \frac{I_0}{(2\pi R)^N}~\tilde\Sigma\,, \ee with $I_0$
representing the Gaussian part. It is now convenient to inverse
Fourier transform the expansion term by term and then evaluate the
Feynman diagrams by summing up the series with exponentials. Thus in
terms of the target space diagrams, the small field expansion for
the anomaly becomes

\ba &&\langle\mbox{Tr}e^{l\phi_{N+1}(\hat x)}\rangle_{\eta_i}
-\langle\mbox{Tr}e^{l\phi_N(\hat x)}\rangle_{\eta_i'}
\nonumber \\
&&= \frac{l}{2\pi R}\Big\langle \frac{1}{\tilde \Sigma}
\mbox{Tr}\Big[e^{l\phi_N(\hat x)}\phi_N^{-1}(\hat x)\Big\{\frac{\pi
R}{M}\coth \pi MR \nonumber \\&&+ \frac{g}{M^4}\int \frac{dt}{2\pi
R}\sum_m\Big(1+\frac{m^2}{R^2M^2}\Big)^{-1}e^{im(\hat x -
t)/R}\phi(t)\sum_n\Big(1+\frac{n^2}{R^2M^2}\Big)^{-1}e^{in(t-\hat
x)/R} \nonumber \\ &&+ \frac{g^2}{2M^4}\int \frac{dt_1 dt_2}{(2\pi
R)^2}\sum_m \Big(1+\frac{m^2}{M^2R^2}\Big)^{-1}e^{im(\hat
x-t_1)/R}\phi(t_1)\sum_n\frac{e^{in(t_1-t_2)/R}}{n^2/R^2+M^2}
\phi(t_2)\nonumber
\\&&\sum_q\Big(1+\frac{q^2}{M^2R^2}\Big)^{-1}e^{iq(t_2-\hat x)/R}
\nonumber \\ &&+ \frac{g^3}{6M^4}\int \frac{dt_1dt_2dt_3}{(2\pi
R)^3}\sum_m \Big(1+\frac{m^2}{M^2R^2}\Big)^{-1}e^{im(\hat
x-t_1)/R}\phi(t_1) \sum_n \frac{e^{in(t_1-t_2)/R}}{n^2/R^2+M^2}
\phi(t_2) \nonumber \\&&\frac{e^{iq(t_2-t_3)/R}}{q^2/R^2+M^2}
\phi(t_3) \sum_k \Big(1+\frac{k^2}{M^2R^2}\Big)^{-1}e^{ik(t_3-\hat
x)/R} +\ldots \Big\}\Big] \Big\rangle_{\eta_i'} \,\,.\ea Now to
evaluate the Feynman diagrams, we use the open string moduli $\hat
x$ to replace the inverse propagator by a differential operator of
$\hat x$

\be \mathcal{O}_{\hat x} \equiv
\Big(1-\frac{1}{M^2}\frac{\partial^2}{\partial\hat
x^2}\Big)^{-1}\,.\label{hatx-o}\ee This  simplifies the summation
over exponentials in the above expression to delta functions
$\delta(t_i-\hat x)$ that forces any matrix quantum mechanics time
$t_i$ (matter) to be at $\hat x$, which is basically reflecting the
fact that we are talking about Dirichlet boundaries that are
introducing explicit dependence on the open string moduli due to
nonperturbative effects coming from all the $N^2$ quantum mechanical
degrees of freedom. Note that replacing the inverse propagator by
(\ref{hatx-o}) is not only consistent formally but also can be seen
to be in agreement as an expansion in the large $M$ limit for
Dirichlet boundaries. In the extreme $M \to \infty$ limit, the
explicit open string moduli dependence is washed out. After
performing the delta function integrals over the matrix quantum
mechanics time, the diagrammatic expansion can now be written as

\ba &&\langle\mbox{Tr}e^{l\phi_{N+1}(\hat x)}\rangle_{\eta_i}
-\langle\mbox{Tr}e^{l\phi_N(\hat x)}\rangle_{\eta_i'}
\nonumber \\
&&= \frac{l}{2\pi R}\Big\langle\frac{1}{\tilde\Sigma}\Big[\frac{\pi
R}{M}\coth \pi MR~\mbox{Tr}(e^{l\phi_N(\hat x)}\phi_N^{-1}(\hat x))
\nonumber \\
&&+ \frac{2\pi R g}{M^4}\mbox{Tr}(e^{l\phi_N(\hat
x)}\phi_N^{-1}(\hat x)~\mathcal{O}_{\hat x}\phi(\hat
x)\mathcal{O}_{\hat x})+\frac{\pi R}{M}\coth\pi
MR~\frac{g^2}{2M^4}~\mbox{Tr}(e^{l\phi_N(\hat x)}\phi_N^{-1}(\hat
x)~\mathcal{O}_{\hat x}\phi^2(\hat x)\mathcal{O}_{\hat x}) \nonumber
\\ &&+\frac{2\pi R g^3}{6M^8}~\mbox{Tr}(e^{l\phi_N(\hat
x)}\phi_N^{-1}(\hat x)\mathcal{O}_{\hat x}\phi(\hat
x)\mathcal{O}_{\hat x}\phi(\hat x)\mathcal{O}_{\hat x}\phi(\hat
x)\mathcal{O}_{\hat x})+....\Big]\Big\rangle_{\eta_i'} \,,\ea The
above diagrammatic expansion can now be arranged according to even
and odd powers of matrix couplings which sum up to even and odd
series of $\frac{g}{M^2} \mathcal{O}_{\hat x}\phi(\hat x)$, thus
computing the anomaly of the form

\be \langle\mbox{Tr}e^{l\phi_{N+1}(\hat x)}\rangle_{\eta_i}
-\langle\mbox{Tr}e^{l\phi_N(\hat x)}\rangle_{\eta_i'}= \frac{l}{2\pi
R}\langle\tilde\Sigma^{-1} ~\mbox{Tr}[(e^{l\phi_N(\hat
x)}\phi_N^{-1}(\hat x))(S_1+S_2)]\rangle \label{anomaly}\,,\ee where
the diagrams with even powers in coupling are given by

\ba S_1 &=& \frac{\pi R}{M} \coth\pi MR
\big[1+\frac{1}{2!}\frac{1}{M^2}~\mathcal{O}_{\hat x}g^2\phi^2(\hat
x)\frac{1}{M^2}\mathcal{O}_{\hat x} \nonumber \\
&&+\frac{1}{4!}\frac{1}{M^2}\mathcal{O}_{\hat x} g\phi(\hat
x)\frac{1}{M^2}\mathcal{O}_{\hat x}g^2\phi^2(\hat
x)\frac{1}{M^2}\mathcal{O}_{\hat x} g\phi(\hat
x)\frac{1}{M^2}\mathcal{O}_{\hat x} + \ldots\big] \nonumber \\
&=& \frac{\pi R}{M}\coth \pi MR ~\cosh
(\frac{g}{M^2}\mathcal{O}_{\hat x}\phi(\hat x))\,, \ea and the odd
powers in coupling are given by

\ba S_2 &=& 2\pi R \big[\frac{1}{M^2}\mathcal{O}_{\hat x} g\phi(\hat
x)\frac{1}{M^2}\mathcal{O}_{\hat x} + \nonumber
\\&&\frac{1}{3!}\frac{1}{M^2}\mathcal{O}_{\hat x}g\phi(\hat x)
~\frac{1}{M^2}\mathcal{O}_{\hat x}g\phi(\hat x)
~\frac{1}{M^2}\mathcal{O}_{\hat x}g\phi(\hat
x)~\frac{1}{M^2}\mathcal{O}_{\hat x} + \ldots\big] \nonumber \\
&=&2\pi R\frac{1}{M^2}\mathcal{O}_{\hat
x}~\sinh(\frac{g}{M^2}\mathcal{O}_{\hat x}\phi(\hat x)) \,.\ea Note
that, adding up the series neatly to hyperbolic functions is
consistent with the small field approximation as well as the large
$M$ limit for the Dirichlet boundaries.

\begin{figure} [ht]
\centering
\input{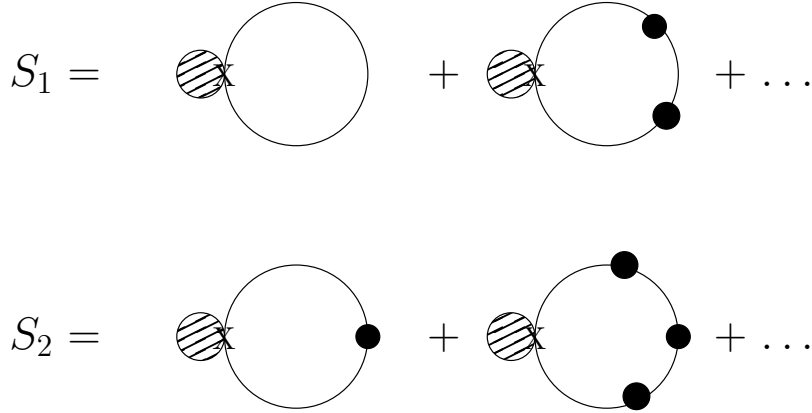} \caption{Diagrams contributing
in $S_1$ and $S_2$.} \label{feynman2}
\end{figure}


\subsection{The modified WdW constraint}
Let us now differentiate both sides of (\ref{anomaly}) with respect
to the length $l$ of the loop to remove the singular $\phi_N^{-1}$
term from the right hand side. Also we do small field expansion of
$\tilde \Sigma ^{-1}$. Let us now recast powers of $\phi_N(\hat x)$
to be multi derivatives of $l$ on the loop operator. This closes the
WdW constraint by factorizing the right hand side of (\ref{anomaly})
into a differential operator of $l$ and $\hat x$ operating on the
one point function of the loop operator and a modified WdW equation
emerges. Thus

\ba \frac{\partial}{\partial
l}\Big[\big\langle\frac{1}{l}\mbox{Tr}~e^{l\phi_{N+1}(\hat
x)}\big\rangle_{\eta_i} -
\big\langle\frac{1}{l}\mbox{Tr}~e^{l\phi_N(\hat
x)}\big\rangle_{\eta_i'}\Big] = && \Big[\frac{1}{2M}~\coth\pi
MR\cosh\Big(\frac{g}{M^2}~\mathcal{O}_{\hat
x}\frac{\partial}{\partial l}\Big) \nonumber \\ &&+
\frac{1}{M^2}~\mathcal{O}_{\hat
x}~\sinh\Big(\frac{g}{M^2}\frac{\partial}{\partial
l}~\mathcal{O}_{\hat x}\Big) \Big]~\langle\mbox{Tr}e^{l\phi(\hat
x)}\rangle_{\eta_i'} + \ldots \,,\nonumber\\\label{fullgenWDW}\ea
where the ellipsis corresponds to the multi-trace or worm hole
deformations that leads to topology changing amplitudes. These are
redundant operators and are in relative $O(1/N)$ compared to the
leading contribution and contribute to topology changing
transitions. Using (\ref{cso}) the linear (dealing with loop of a
single topology) generalized WdW equation can be written as

\ba &&\Big[\Big(-\frac{1}{l^2}+\frac{1}{l}\frac{\partial}{\partial
l}\Big)\Big(N\frac{\partial}{\partial
N}-\beta_{\eta_i}\frac{\partial}{\partial\eta_i}+1\Big)-\frac{N}{2M}\coth\pi
MR\cosh\Big(\frac{g}{M^2}\mathcal{O}_{\hat x}
\frac{\partial}{\partial l}\Big)\nonumber
\\&&-\frac{N}{M^2}\mathcal{O}_{\hat x}\sinh\Big(\frac{g}{M^2}
\frac{\partial}{\partial l} \mathcal{O}_{\hat
x}\Big)\Big]\frac{1}{N}\langle\mbox{Tr}~e^{l\phi_N(\hat
x)}\rangle_{\eta_i'}= 0\,.\label{generalWDW} \ea Clearly the WdW
constraint is {\it modified} or {\it generalized}, as compared to
the usual one, in the sense of being beyond minisuperspace (see the
discussion in section $2$). It captures effects of derivatives of
all orders, which is more than the effect caught by a perturbative
superspace (considering fluctuations on the Fermi surface in the
free fermion language). Also the higher derivative effect captured
here is for all genus wave function as compared to the genus zero
result obtained in the perturbative superspace. Moreover, the scale
factor $l$ here contains all the modes of the metric fluctuation as
compared to the presence of zero mode $l_0$ only in the dynamical
Fermi sea picture. Note the difference in the $l$ dependence of the
coefficients of $V(\partial/\partial l_0)$ and $V'(\partial/\partial
l_0)$ in (\ref{higher-derivative}) and that of the $\cosh$ and
$\sinh$ terms in (\ref{generalWDW}). This point needs to be
understood. Another aspect of (\ref{generalWDW}) being beyond
minisuperspace is that, it has an in built way of determining the
factor ordering ambiguity. However, as we mentioned above, the
linear modified WdW constraint that governs the dynamics of loops of
a particular topology, comes from the leading order contribution of
$\tilde \Sigma^{-1}$. The sub-leading orders, beginning from a
relative $O(1/N)$ compared to the single trace terms, give rise to
the nonlinear part of the WdW constraint corresponds to dynamics of
topological transitions between loops of different topologies. This
is strictly a beyond minisuperspace process that can no way be
captured within minisuperspace. We will return to this in the last
section.

From now on let us write (\ref{generalWDW}) in terms of the wave
function (\ref{wavefn}). In order to reduce the dimension of the
space of independent variables, let us consider the following ansatz
for the Callan-Symanzik equation of the macroscopic loop variables

\be \big(N \frac{\partial}{\partial
N}-\beta_{\eta_i}\frac{\partial}{\partial \eta_i}+1\big)
\psi(l;N,\eta_i,R )=\big(l^2
\mc{F}_1(\eta_i,N,R)+\mc{F}_2(\eta_i,N,R)\big)~\psi(l;N,
\eta_i,R)\,.\label{CS-eigenvaluecond} \ee Here,
$[\mc{F}_1]=[l^{-2}]$ and $\mc{F}_2$ is dimensionless. The ansatz is
chosen keeping in mind that the wave function behaves somewhat like
the exponential of the Euclidean action, the area $l^2$. From our
discussion in the previous subsections we already know that the
macroscopic loop has an anomaly in the Callan-Symanzik equation
which demands that they are annihilated by the WdW operator that
acts as a gauge fixing condition for them. Here, by considering this
ansatz, we are actually assuming that we are looking at those
macroscopic loops which not only have WdW constraint as a gauge
fixing condition, also they themselves are invariant under scale
changes. So that, for them the anomaly basically simplifies into
some kind of {\it anomalous dimension} (a number). This reduces the
space of independent variables to the dynamical variables only, the
length of the loop $l$. As we mentioned before, due to
nonperturbative effect coming from all the $N^2$ quantum mechanical
degrees of freedom, the wave function has explicit dependence on the
open string modulus $\hat x$ which acts as another dynamical
variable. Now our job is to plug in (\ref{CS-eigenvaluecond}) into
(\ref{generalWDW}) and solve for the wave function that as well
determines $\mc{F}_1$ and $\mc{F}_2$ in a self consistent way. The
functions $\mc{F}_1$ and $\mc{F}_2$ keep the information about the
$c=1$ fixed point of the matrix quantum mechanics around which the
solution is being studied.

Let us consider another simplification. Since the time direction is
compact, $\hat x=0$ can be identified with $\hat x= 2\pi R$.
Therefore one can assume the $\hat x$ dependence of the wave
function $\psi(l,\hat x)$ to be periodic. The simplest choice, that
is nonzero at $\hat x=0$, is plane waves $\psi(\hat x)=\cos(
\frac{n}{R} \hat x)$ with quantized momenta. Thus,

\be \mc{O}_{\hat x} \psi(\hat x)=\lambda_n~\psi(\hat
x)\,,~~~~\lambda_n=\frac{M^2}{M^2+n^2/R^2}\,.\label{hatx-eigenveqn}\ee
Thus considering (\ref{CS-eigenvaluecond}), (\ref{generalWDW}) and
(\ref{hatx-eigenveqn}), the beyond minisuperspace WdW equation for
the one point function of loop operator becomes

\be \Big[2l~A \cosh\big(Q~\lambda_n \frac{d}{d l}\big)+ 2l~ B
\lambda_n \sinh\big(Q~\lambda_n \frac{d}{d l}
\big)-(l^2\mc{F}_1+\mc{F}_2) \frac{d}{d l}-2l
\mc{F}_1\Big]\psi(l)=0\,, \label{opfn-wdw} \ee where, \be
A=\frac{N}{2M}\coth\pi
MR\,,~~B=\frac{N}{M^2}\,,~~Q=\frac{g}{M^2}\,.\ee It is important to
note here that all the parameters A,B,Q have $M \to -M$ symmetry
(see the discussion under $c=1$ fixed point in the previous
section).

To study the solution of (\ref{opfn-wdw}) at different scales,
especially at small and large scale limits, it will be useful to
change the dependent variable to one point function with one
puncture

\be \Psi(l,\hat x)=\frac{\psi(l, \hat x)}{l}=\frac{1}{N}\langle
\mbox{Tr}\frac{e^{l \phi_N(\hat x)}}{l} \rangle_{\eta_i'}\,. \ee
This basically defines a closed Dirichlet boundary of length $l$
with one marked point on it. Using the following identity

\be e^{\pm Q \frac{d}{d l}}~\frac{\psi(l)}{l}=\frac{1}{l} ~e^{\pm Q
\frac{d}{d l}} ~\psi(l)+\frac{\psi(l)}{l \pm Q} \ee in
(\ref{opfn-wdw}), the WdW equation for the one point function with
one puncture can be rewritten as

\be \Big[2l~A \cosh\big(Q~\lambda_n \frac{d}{d l}\big)+ 2l~ B
\lambda_n \sinh\big(Q~\lambda_n \frac{d}{d l}
\big)-(l^2\mc{F}_1+\mc{F}_2) \frac{d}{d l}-2l \mc{F}_1-\frac{2l^2(A
l-B Q \lambda_n^2)}{l^2-Q ^2\lambda_n^2}\Big]\Psi(l)=0
\label{onepuncture-opfn-wdw} \,.\ee This actually induces a
polynomial potential term in the WdW equation that brings in the
notion of a special scale $\tilde l(n)=Q \lambda_n$ separating the
$UV$ and $IR$ physics. From $2D$ cosmology point of view this is
naturally a inverse Hubble scale $H^{-1}$ for the problem. In the
next section we will discuss the significance of this scale in
detail and study physics of the solution to
(\ref{onepuncture-opfn-wdw}) with respect to it.


\section{The cosmological implications of the wave function}
\setcounter{equation}{0}

In the previous section we have seen that the modified WdW equation
(\ref{onepuncture-opfn-wdw}) for one point function with one
puncture has a polynomial potential that naturally brings in the
notion of a special scale $\tilde l(n)= Q \lambda_n$ for the
problem. In this section we will study the cosmological implication
of this scale and the study the small and large scale behavior of
the solution to (\ref{onepuncture-opfn-wdw}) with respect to this
scale.


\subsection{The inverse Hubble scale}

Let us consider a rough qualitative analysis of the shape of the
polynomial potential in (\ref{onepuncture-opfn-wdw}) with respect to
the scale $l$. Here one can expand the hyperbolic functions of
differential operator and write down the potential with respect to
the quadratic differential term  $(-l^2 \frac{\partial^2}{\partial
l^2})$ in the Liouville metric $d \varphi=dl/l$. Such an expansion
is possible considering $Q=g/M^2$ to be small, which is definitely
the case around a $c=1$ fixed point that has small $g$ and for
Dirichlet boundary that has large $M$. The potential is given by

\be V(l,n)=\frac{2l^2}{\tilde l(n)^2}
\Big(\frac{\mc{F}_1}{A}+\frac{l(l-\frac{B \lambda_n}{A}\tilde
l(n))}{(l^2 -\tilde l(n)^2)} +1\Big)\,.\label{potential}\ee Here $n$
is related to the momentum modes of matter. Since
$[\mc{F}_1]=[A]=[l^{-2}]$, the ratio $\mc{F}_1/A$ is purely a
number. Now the structure of the potential depends on the size of
the quantity $B \lambda_n/A$. For Dirichlet boundaries

\be B \lambda_n/A= 2 M \tanh(\pi M R)/(M^2+\frac{n^2}{R^2}) \sim
2/M<<1\,.\label{mqm-param-constraint}\ee This shows that for
$l<<\tilde l(n)$, the potential $V(l,n)$ looks like a barrier in a
tunneling problem.

This is more clear from the form of the potential (\ref{potential})
along with the constraint (\ref{mqm-param-constraint}) in $\hat x$
space

\be V(l,\hat x)=V_{local}+ \frac{g^2}{l^2} \frac{\pi
R}{\sqrt{M^2-g^2/l^2}} \frac{\cosh \sqrt{M^2-g^2/l^2}(\pi R-\hat
x)}{\sinh \sqrt{M^2-g^2/l^2}~\pi R}(2-4 \pi \tanh \pi M R l/g)\,,\ee
where

\be V_{local}=\frac{2 l^2}{g^2}[2-(C^{-1}+2 C e^{-C})][M^2
\delta(\hat x)+\delta''(\hat x)]+(2-4 M \tanh \pi M R
l/g)\delta(\hat x)\,,\ee being proportional to $\delta(\hat x)$ can
be neglected by considering $\hat x \ne 0$. Thus the magnitude of
the matrix quantum mechanics parameters around the $c=1$ fixed point
as compared to the smallness or largeness of $l$ and the form of
$\tilde V= V- V_{local}$ controls the behavior. Let us consider
that, as $l \to 0$, $g/l \ll M$. Then, assuming $\hat x \sim \pi R
\sim 1$, $\tilde V$ simplifies to

\be \tilde V(l \to 0) \sim \Big (M e^{-M} \big(\frac{2 g^2}{l^2
M^2}\big)-4 \pi e^{-M}\frac{g}{l M} \Big)\to 0\,. \ee On the other
hand $\tilde V (l \to \infty) \to -\infty$ and $\tilde V \to 0$ at a
scale $l \to O(g/M) > \tilde l \sim O(g/M^2)$. This provides the
tunneling barrier. Recall that the special scale $\tilde l $ occurs
before the potential goes to zero.

This indicates that for small scale $l<\tilde l$, the $1D$ closed
universe with homogenous matter that we are considering, presumably
admits a tunneling type of wave function that creates the universe
from nothing by Euclidean tunneling. In cosmological models of $4D$
gravity such tunneling wave function have been shown to predict
inflation {\cite{v-qcis}}. Whether the tunneling actually predicts
an inflationary scenario depends on the scale to which universe
tunnels to. If the scale expands exponentially in a short time, then
the universe undergoes inflation. At large scale $l>\tilde l$ the
potential approaches $-\infty$ and hence the wave function of the
universe presumably admits outgoing plane wave that gives an
expanding de Sitter universe in the far future. In the following
subsections we will show that the phase $l<\tilde l$ will be the
phase at which universe undergoes inflation and the phase $l>\tilde
l$ will be the far future phase of an expanding universe which is
already hugely inflated. The region, where $l=\tilde l$, actually
gives the hot big bang.

Thus the scale $\tilde l(n)$ acts like an inverse Hubble scale
$H^{-1}$ that separates the small and large scale behavior of the
wave function. For $c=1$ matrix model, it is at very short distance
as would be appropriate for an inflationary scenario. Note that the
nonperturbative effects due to the $N^2$ quantum mechanical degrees
of freedom is introducing explicit open string moduli dependence of
the Hubble scale. Though the moduli dependence does not directly
affect the inflationary scenario, as one always has a tunneling wave
function solution for any sensible constant $H$, its explicit
dependence determines the geometry of the vacuum one tunnels to. The
matter dependence of this scale can be written down by inverse
Fourier transforming $\tilde l(n)$ as

\be \tilde l(\hat x)=H(\hat x)^{-1}=\frac{\pi g R}{M}
\frac{\cosh(\pi M R-M \hat x)}{\sinh(\pi M R)} \,,~~~~0 \le \hat x
\le 2 \pi R\,.\label{Hubble-scale}\ee This is the scale factor
corresponding to a de Sitter geometry of the world sheet

\be ds^2=dX^2-\frac{g^2 R^2}{M^2} \cosh^2(\pi M R-M X)
~d\varphi^2\,.\ee Recall that the world sheet matter $X(\sigma)$ is
mapped to the matrix quantum mechanics time through the delta
function in the Dirichlet boundary condition (\ref{dbc}). The issue
of the Lorentzian signature is related to the fact that the large
$N$ RG with $N^2$ matrix quantum mechanics degrees of freedom
captures the de Sitter sign of the cosmological constant ($\mu<0$)
in the WdW equation (like the supercritical case) instead of the
usual $AdS$ sign ($\mu>0$) of the $c=1$ quantum mechanics. This
effectively implies an imaginary slope of linear dilaton
($\gamma^2<0\,, \gamma \to i \gamma\,,Q=\gamma/2 \to iQ$) and hence
indirectly implies growth of number of space-time dimensions. On the
other hand the exponential potential wall $e^{\gamma \varphi}$
remains the same in the WdW equation. Hence with $\gamma \to i
\gamma$, the Liouville field also becomes imaginary $\varphi \to -i
\varphi$. This takes care of the reality of the Liouville action and
gives rise to a Lorentzian metric.


\subsection{The small scale behavior: inflation}

Let us now solve for the small scale behavior of the wave function.
For $l<\tilde l=Q \lambda_n$, The polynomial term can be expanded as

\be P(l)=\frac{2 l^2}{\tilde l^2} (A l-B \lambda_n \tilde l)\sum
_{m=0}^{\infty} \frac{l^{2m}}{\tilde l^{2m}} \label{polynomial}\,.
\ee As indicated by the potential (\ref{potential}), the wave
function in the region $l<\tilde l$ can have tunneling through the
barrier. Thus an appropriate trial solution satisfying
(\ref{onepuncture-opfn-wdw}) would be

\be \Psi(l) \sim a(N, \eta_i) e^{-\Delta(\eta_i) l^2}
\label{trial-soln} \ee Later we will argue that such a choice of
trial function is a unique choice fixed by the initial profile of
the scale fluctuation $\Psi(0+\delta l)$. Now using the identities

\ba \cosh(\tilde l \frac{d}{dl}) e^{-\Delta l^2}=\cosh(2 \Delta
\tilde l l)
e^{-\Delta (l^2+\tilde l^2)}\,,\nonumber\\
\sinh(\tilde l \frac{d}{dl}) e^{- \Delta l^2}=\sinh(2 \Delta \tilde
l l) e^{-\Delta (l^2+ \tilde l^2)} \,,\ea and expanding the
hyperbolic functions in power series of $l$ (for $\tilde l <1$ {\it
i.e.} $g < M^2$ and for a small $l$), we set the coefficients of all
the powers of $l$ to be zero. Thus we see that $\Psi(l)\sim a(N,
\eta_i) e^{-\Delta (\eta_i)l^2}$ is a solution of
(\ref{onepuncture-opfn-wdw}) with the spectrum $\Delta$ determined
by

\ba &&e^{-\Delta \tilde l^2}+ \frac{\Delta
\mc{F}_2}{A}-\frac{\mc{F}_1}{A}=0\,,
\nonumber\\
&& \Delta^2~\tilde l^2~e^{-\Delta \tilde l^2}+\frac{\Delta
\mc{F}_1}{2 A}+\frac{1}{2 \tilde l^2}=0 \nonumber\\
&&\frac{(2 \Delta \tilde l)^n}{n!} e^{-\Delta \tilde l^2}+(-1)^n
\frac{1}{\tilde l^n}=0\,,~~n \ne 0,2\,. \label{delta-eqn}\ea Here we
have three equations and three unknowns $\Delta, \mc{F}_1 $ and
$\mc{F}_2$. The equations (\ref{delta-eqn}) indicate that any
solution for the spectrum $\Delta$ should go somewhat like
Lambert$W$ function. For Lambert$W$ functions the function itself is
in the same footing as it's exponential. Hence such a solution for
the spectrum $\Delta$ will be a good candidate to actually perform
the cancelation of the hyperbolic function by a polynomial
(\ref{polynomial}) needed to  satisfy equation
(\ref{onepuncture-opfn-wdw}). Thus solution to the spectrum $\Delta$
can be written as $\Delta=C/\tilde l^2$, where the number $C$ is
given by the three sets of relations in combined form

\be e^{-C}=\frac{n!(-1)^{n+1}}{(2C)^n}
=-\Big(\frac{1}{2C^2}+\frac{\mc{F}_1}{2A C}\Big)
=\frac{\mc{F}_1}{A}-\frac{C \mc{F}_2}{A \tilde l^2}\,,~~~~n \ne
0,2\,.\ee Now solving these relations we have \ba &&C=-n~
\mbox{LambertW}\Big(\frac{(-n!)^{1/n}}{2
n}\Big)\,,~~~n \ne 0,2\,,\nonumber\\
&&\mc{F}_1=A \Big(-\frac{1}{C}-2C e^{-C}\Big)\,,\nonumber\\
&&\mc{F}_2=A ~\tilde l^2 ~\Big(
-\frac{1}{C^2}-\frac{e^{-C}}{C}-2e^{-C}\Big)\,.\label{spectrum-wfn}\ea
Thus, $\Delta$ is given by

\be \Delta = -\frac{n}{g^2}~
\mbox{LambertW}\Big(\frac{(-n!)^{1/n}}{2
n}\Big)\Big(M^2+\frac{n^2}{R^2}\Big)^2\,,~~~n \ne
0,2\,.\label{delta-spectrum}\ee

In order to give an idea, the table below illustrates different
values of $C$ corresponding to sample arguments, using
(\ref{spectrum-wfn}).


\vskip0.5cm

\begin{tabular}{|c|c|}
\hline n & C \\
  \hline
1 & -0.79+0.77i \\
3 & 1.5 \\
4 & 0.62+1.13i \\
5 & 1.91 \\
6 & 1.4 +1.27i \\
7 & 2.37 \\
8 & 1.49+0.8i \\
9 & 2.84 \\
10 & 2.68+1.26i \\
11 & 3.69 \\
  \hline
\end{tabular}

\vskip0.5cm


However, from (\ref{CS-eigenvaluecond}) $\Delta, \mc{F}_1, \mc{F}_2$
and the amplitude of the wave function $a$ satisfy additional
constraints

\ba && \beta_{\eta_i} \frac{\partial \Delta}{\partial \eta_i}
=\mc{F}_1\,,\nonumber\\
&&~N \frac{\partial a}{\partial N}-\beta_{\eta_i}\frac{\partial
a}{\partial \eta_i}+a=\mc{F}_2 a\,.\label{addnl-constraint}\ea In
writing down this constraints we considered the fact that all the
four parameters $\Delta, \mc{F}_1, \mc{F}_2, a$ are independent of
$l$ and $\Delta$ does not have any explicit $N$ dependence as
indicated by the solution (\ref{delta-spectrum}). Obviously the
second one of the two relations in (\ref{addnl-constraint})
determines the unknown parameter $a$. Now we need to check whether
the other one is really a redundant condition or a serious
constraint on the already determined parameters. To check this, let
us recall that around $c=1$ fixed point we have $h=0$, {\it i.e.}
$\beta_g \sim -(g-g^*)/2$ and $\beta_{M^2} \sim 0$. Thus, as $g \to
g^*$ around a $c=1$ fixed point, solving (\ref{addnl-constraint})
for the ansatz $\Delta= \mc{F}_1 \Delta(g)$ gives

\be \Delta(g)=\frac{K}{(g-g^*)^2}\Rightarrow \Delta= \frac{\mc{F}_1
K}{(g-g^*)^2} \,. \label{delta-addnlc}\ee Here $K$ is nothing but an
undetermined integration constant. Choosing this constant to be $K=C
(M^2+\frac{n^2}{R^2})^2/\mc{F}_1$ makes the solution for $\Delta$ in
(\ref{delta-addnlc}) equivalent to that already given by
(\ref{delta-spectrum}). Thus the first of the two relations in
(\ref{addnl-constraint}) is a redundant constraint for $\Delta$. Now
let us solve the second relation in (\ref{addnl-constraint}) by
separation of variables, $a=\frac{1}{N}~\tilde a(g,M,R)$, which (up
to a constant factor) gives

\be a \sim \frac{1}{N}~(g-g^*)^{\frac{N g*^2}{M g^2}\coth(\pi M
R)~\Big( -\frac{1}{C^2}-\frac{e^{-C}}{C}-2e^{-C}\Big)~\tilde l^2}~
e^{\frac{N}{M}(\frac{1}{2}+\frac{g*}{g})\coth(\pi M R)~\Big(
-\frac{1}{C^2}-\frac{e^{-C}}{C}-2e^{-C}\Big)~\tilde l^2}\,.\ee Here
$C$ is given by the spectrum (\ref{spectrum-wfn}). We see that (in
accordance with \cite{bdss}) the multiplicative prefactor $a$
absorbs all the non-universal parameter dependence of the wave
function and the universal part goes like $\Psi(l<\tilde l) \sim
e^{-C\frac{l^2}{\tilde l^2}}$. Note that, the $R$ dependence in the
$\tilde l$ comes only through the Fourier modes of matter indicating
{\it matter dependence}. Thus the full solution for the wave
function at short distance ({\it i.e.} below Hubble scale) can be
summarized to be

\ba &&\Psi(l<\tilde l)=\frac{1}{N}~(g-g^*)^{\frac{N g*^2}{M
g^2}\coth(\pi M R)~\Big(
\frac{1}{C^2}-\frac{e^{-C}}{C}-e^{-C}\Big)~\tilde
l^2}\nonumber\\&&~~~~~~\times e^{\frac{N (g+g*)^2}{2 M g^2}\coth(\pi
M R)~\Big( \frac{1}{C^2}-\frac{e^{-C}}{C}-e^{-C}\Big)~\tilde
l^2}~e^{-C~l^2/\tilde l ^2}\,, \label{ss-wfn}\ea where $C$ is given
by (\ref{spectrum-wfn}). Note that the nucleation probability given
by this wave function is a mixture of the usual exponential
suppression and the power law $\mu^{\delta}$ (here $\mu \sim
(g-g^*)$), exhibited by the inflating lower energy phase in the
critical droplet fluctuation in $2D$ Liouville gravity, that tends
to mimic the behavior of dynamical lattice Ising model
\cite{zz-dmv}.

The first few values of $C$ computed in table above shows that for
$n \ge 3$ the wave function indeed is a decaying wave function. In
this regime, for odd $n$, $C \ge 0$ and the wave function is a
purely decaying wave function. For even $n$, $C$ is complex. Hence
the wave function has a outgoing plane wave part along with a
decaying part. Note that, only for $n=1$, the wave function has a
growing behavior (the usual Hartle-Hawking behavior) instead of a
decaying one, along with an outgoing plane wave. However, $n$ here
are in one to one correspondence with the orders of the derivatives
of $l$. For computations away from minisuperspace, indeed $n \ge 3$,
and hence there are enough spectra giving a decaying solution. The
growing solution can be suppressed by the choice of the boundary
condition $\Psi(0) =\Psi_0 \ne 0$. Another point to be noted here is
that, since for any value of $n$ the wave function does not have any
incoming plane wave part, one can say that the closed universes
described by our wave function never re-collapses.

Such an exponentially decaying wave function can be interpreted to
describe quantum creation of a closed inflationary universe filled
with homogenous field $\hat x$ tunneling from a state of vanishing
scale or {\it nothing}. To understand what kind of vacuum it tunnels
to, let us consider the functional dependence of $\tilde l$ on $\hat
x$ from (\ref{Hubble-scale}). At the end of the tunneling,
eventually at a scale $\tilde l(\hat x)$, the universe in a state of
large total energy (for example note that, in the standard chaotic
inflation scenario, the initial energy density is of the order of
$M^4_p$) $E \sim M$ (here $[M]=[l^2]=[\hat x^{-1}]=[E]$) will
undergo a huge exponential expansion over a very short period of MQM
time $M^{-1}$ or due to a very small quantum fluctuation $\delta t
\sim M^{-1}$. This is indeed happening for Dirichlet boundaries
($\delta \hat x \sim O(1/M)\to 0$). At the Hubble scale, to which
the universe tunnels to, a very small quantum fluctuation sets an
exponential growth in scale for a universe of huge energy. {\it Thus
for $l < \tilde l$, (\ref{ss-wfn}) is the quantum wave function
representing an one dimensional homogenous universe tunneling from
`nothing' to a state of chaotic inflation.}

Note that the Hartle-Hawking wave function computed away from the
minisuperspace approximation, predicts `tunneling to an inflationary
vacuum from nothing', in a structurally similar way (apart from the
fact that here $l=l(\phi(\tau))$ and not just $l=l_0(t)$) to the
Linde-Vilenkin tunneling wave function in minisuperspace
cosmological model

\be \Psi(l_0<H^{-1}) \sim e^{-\frac{\pi}{2}H^2 l_0^2}\,.\ee This is
unlike the usual growing behavior of the Hartle-Hawking wave
function that makes predicting inflation to be impossible. One can
argue that the boundary condition $\Psi(0)=\Psi_0 \ne 0$ chosen here
may introduce arbitrariness in the initial condition of the
universe. However, the Hartle-Hawking wave function $\Psi(l)$ is the
wave function for the one point function with one puncture. The wave
function for the $1D$ universes, in the true sense of the term (or
according to traditional definition of the term), is the smooth one
point function $\psi(l)=l~\Psi(l)$ of the loop operator. This
vanishes in the UV and hence has no arbitrariness in the initial
condition. It has a growing and then rapidly decaying behavior
capturing inflation.

Let us now argue for the uniqueness of the choice of our trial
solution (\ref{trial-soln}). As the WdW equation
(\ref{onepuncture-opfn-wdw}) is of infinite degree, in principle one
needs an infinite dimensional initial condition to uniquely choose
the solution. Here we argue that a gaussian profile of the initial
($l=0$) scale fluctuation

\be \Psi(0+\delta l)=\Psi_0 e^{-\Delta \delta l^2} \,,\ee fixes the
initial values of all the derivatives

\be
\partial^{2n+1}_l\Psi(0)=0\,,
~~~\partial^{2n}_l\Psi(0)=\frac{(2n)! \Delta^n}{n! (-1)^n}\Psi_0\,.
\ee This provides the infinite dimensional initial condition to
uniquely choose the wave function of the form $\Psi(l< \tilde l)
\sim e^{-\Delta l^2}$ as the solution of
(\ref{onepuncture-opfn-wdw}).

Let us now explain the significance of the gaussian profile of the
initial scale fluctuation. The density perturbation due to the scale
fluctuation of the one dimensional universe we are considering, is
given by

\be \frac{\delta \rho}{\rho} \sim \frac{\delta l}{l}=P_{l \to
l+\delta l}=\frac{\Psi(l+\delta l)}{\Psi(l)}\,,
\label{density-pertb} \ee where we define $P_{l \to l+\delta l}$ to
be the probability of finding an universe of scale $l$ between $l$
and $l+\delta l$. As we know, in an exponentially expanding
universe, $H^{-1}$ acts as a cut-off wavelength. The vacuum
fluctuations with a wavelength greater than $H^{-1}$ freezes as a
classical field, producing density perturbations \cite{linde-ppic,
linde-isc}. Now using (\ref{ss-wfn}) in (\ref{density-pertb})

\be \frac{\delta l}{l}=e^{-C\frac{l^2}{\lambda^2}[2 \frac{\delta
l}{l}+(\frac{\delta l}{l})^2]} \simeq \big(1-2C l^2/\lambda^2~\delta
l/l  +O(\delta l/l)^2 \big)\,, \ee where $\lambda=H^{-1}$. This
gives

\be \frac{\delta l}{l}\sim \frac{\lambda^2}{\lambda^2+2 C l^2}
\,.\ee Computing the quantity (\ref{density-pertb}) for $l \gtrsim
\lambda=H^{-1}$ we have

\be  \frac{\delta \rho}{\rho} \sim \frac{\delta l}{l}\simeq
\frac{\lambda^2}{2 C l^2}+O(\lambda^4/l^4)\,.\ee This implies a {\it
nearly nonflat} spectrum of density fluctuation at $l \gtrsim
\lambda=H^{-1}$({\it i.e.} at big bang). For example, for $\lambda
\sim l^{1.01}$ (recall $l<<1$ at the initial moment), in order to
achieve this bound on the wavelength,

\be \frac{\delta \rho}{\rho} \sim l^{0.02}\,.\ee {\it This implies a
scalar spectral index $n_s =0.96$.} Thus, presumably one could argue
that the gaussian initial scale fluctuation profile that uniquely
chooses the wave function is not an adhoc assumption, rather it
physically arises from a special scale dependence of the density
perturbations.


\subsection{The large scale behavior: far future de Sitter}

Let us now consider the large scale limit $l> \tilde l$. We would
like to know whether the wave function in this region behaves like
outgoing plane wave. This would be interesting cosmologically, as
outgoing (incoming) plane waves in this region would represent
expanding (contracting) global De-Sitter geometry in the far future
\cite{dcm}. However, universe at large scale is a classical object.
So the semiclassical wave function extrapolated to far future is
able to capture such expanding geometries. The inability of the
semiclassical wave function is in capturing the inhomogeneities that
sets in due to quantum fluctuation in the early universe.

In the large scale limit $l>\tilde l$, the WdW equation
(\ref{onepuncture-opfn-wdw}) can be written as

\be \Big[A
 \cosh\big(\tilde l \frac{d}{d l}\big)+ B \lambda_n
 \sinh\big(\tilde l \frac{d}{d l}
\big)-\frac{\mc{F}}{2}\frac{d}{dl}-A\Big]\Psi(l)=0
\,.\label{l-large}\ee Here we have rewritten
(\ref{CS-eigenvaluecond}) appropriately for the wave function
proportional to a plane wave in $l$ space and kept only the
contribution from the anomalous dimension that is significant in the
large scale limit. Clearly the wave function at large scale can
assume (outgoing and incoming) plane wave solutions of the form
$\Psi(l)=e^{\pm i \Delta l}$ with

\be \mc{F}=N \frac{\partial \Delta}{\partial
N}+\frac{g}{2}\frac{\partial \Delta}{\partial g}=0\ee Here we have
used the fact that around $c=1$ fixed point, the betafunctions
behaves as $\beta_g \sim -g/2$, $\beta_{M^2} \sim 0$, with $N g^2
=1/g_{st}=const$. This forces $\mc{F}$ to vanish automatically. The
corresponding spectrum of Dirichlet boundaries at large scale (the
one dimensional universes with homogenous field) is then given by

\be A \cosh(i \tilde l ~\Delta)\pm B \lambda_n \sinh(i \tilde l~
\Delta)-A=0\,. \ee This implies the wave function at large scale is
an outgoing plane wave (or a far future expanding De-Sitter
geometry) with the spectrum \be \Delta_{m,n}=\frac{2 m \pi}{\tilde
l}=\frac{2m \pi}{g}(M^2+n^2/R^2)\,,~~~m,n=0, \pm 1,\pm 2, \ldots \ee
Again, the non-universal part of the spectrum is only due to Fourier
modes of matter and factorizes in the wave function. The universal
part of the spectrum goes like $\Delta_m = \frac{2m\pi M^2}{g}$.


\subsection{The de Sitter minisuperspace?}

Here we will show that suppressing the effect of higher order
derivatives with respect to the scale in
(\ref{onepuncture-opfn-wdw}) leads to a de Sitter minisuperspace
type of WdW cosmology, namely, a global de Sitter in the far past
instead of a tunneling wave function for an inflationary universe.
The far future dynamics of global de Sitter is unchanged, as would
be expected.

For this purpose, let us consider the expansion of the hyperbolic
differential operators in (\ref{onepuncture-opfn-wdw}) using $\tilde
l$ as the expansion parameter. This is consistent as $\tilde l<<1$
around a $c=1$ fixed point. We also consider $l \lesssim \tilde l$
and $B/A \sim O(1/M) \to 0$. The denominator of the fractional
polynomial in $l$ in the equation (\ref{onepuncture-opfn-wdw}) can
be expanded in powers of $l/\tilde l$. Thus keeping up to
$O(l/\tilde l)^2$ and second order derivatives in $l$,
(\ref{onepuncture-opfn-wdw}) can be rearranged as

\be \Big[-l^2 \frac{d^2 }{d l^2}+\frac{\mc{F}_2 }{A \tilde l^2}~ l
\frac{d }{d l}-\frac{2}{\tilde l^2} \Big(1-\frac{\mc{F}_1}{A}
\Big)l^2\Big] \Psi(l)=0\,.\label{l-minisup}\ee This is structurally
very similar to a the minisuperspace WdW equation. To make contact
with all genus minisuperspace WdW equation, one can in principle
allow more terms in the potential, say $l^4$ term,  by truncating
the fractional polynomial potential in (\ref{onepuncture-opfn-wdw})
at orders higher than $O(l/\tilde l)^2$. We identify the factor
ordering ambiguity $p$ and the world sheet cosmological constant
$\mu$ to be

\ba && p=-\frac{\mc{F}_2 }{A \tilde
l^2}=\Big(2e^{-C}+\frac{e^{-C}}{C} +\frac{1}{C^2}\Big)\,,\nonumber\\
&& 4\mu=-\frac{2}{\tilde l^2} \Big(1-\frac{\mc{F}_1}{A}
\Big)=-\frac{2}{\tilde l^2} \Big(1+2C e^{-C}+\frac{1}{C}\Big)<0\,,
\label{mu-p}\ea with $C$ given by (\ref{spectrum-wfn}). However, the
De-Sitter sign of the cosmological constant $4 \mu <0$ clearly shows
that the WdW equation derived with all modes of $\varphi$ turned on
({\it i.e.} derived with $N^2$ quantum mechanical degrees of
freedom) but truncated to space with quadratic scale derivatives is
a De-Sitter minisuperspace WdW equation like that of a supercritical
theory. This is different from the usual $AdS$ minisuperspace WdW
equation (\ref{minisupwdw}) or (\ref{ads-qm-f}) of the (non)critical
theory derived with zero mode truncation of dilaton $\varphi_0$
({\it i.e.} with $N$ quantum mechanical degrees of freedom) in a
quadratic space of scale derivatives. This indirectly indicates a
growth in space-time directions. The wave function given by
(\ref{l-minisup}) behaves as the modified Bessel function \be
\Psi(l) \sim l^{\frac{1}{2}(p+1)} J_{-\frac{1}{2}(p+1)}(2 \sqrt{-
\mu}l)\,, \ee and thus would go like the De-Sitter minisuperspace
wave function $J_0(\sqrt{-\mu}l)$ at $p=-1$. This represents a
global de Sitter in the far past \cite{dcm}.

Recall that the factor ordering ambiguity in the minisuperspace WdW
equation (\ref{minisupwdw}) in Liouville background is $p=-1$. This
makes the wave function purely Bessel function and behaves as plane
wave at $l \to 0$ rather than a tunneling wave function. Actually
the solutions to (\ref{minisupwdw}) are different Bessel or Hankel
functions depending on appropriate boundary conditions and the sign
of $\mu$.  They give rise to different classical geometries
corresponding to homogenous FRW universes. For example, Bessel
$K_0(2\sqrt{-\mu}l)$ gives rise to AdS geometry ($\mu>0$)
corresponding to big bang and big crunch while Bessel
$J_0(2\sqrt{-\mu}l)$ and Hankel $H_0({2\sqrt{-\mu}}l)$ can be shown
to give global de Sitter geometry ($\mu<0$) in the far past and far
future respectively \cite{dcm}. These solutions are unable to
capture any kind of inflationary scenario that introduces large
inhomogeneities with evolution of time. In our large $N$ RG, we see
that we have a more general factor ordering ambiguity that can admit
other values too. For $p=-1$ we recover the structure of the usual
minisuperspace WdW in MQM.

However, the quantum wave function in the $l> \tilde l$ regime for
Dirichlet boundaries has similar behavior as the minisuperspace wave
function extrapolated to the large scale. For example the $l> \tilde
l$ solution of (\ref{l-large}), outgoing plane waves characterizing
a far future de Sitter, can be obtained from the large scale
extrapolation of the minisuperspace wave function, a Bessel
function. Thus the far future solutions for universe with homogenous
field can be qualitatively read from the minisuperspace wave
function. presumably this is because the inhomogeneities at the
large scale are captured more successfully by the dynamics of the
nonzero modes of matter that is not dealt by the Dirichlet
boundaries. Perhaps one needs to learn this dynamics from Neumann
boundaries ($D$-1 branes) in $2D$ string theory.

Let us make a brief comment here regarding the magnitude of the
de-Sitter cosmological constant as captured by large $N$ RG. From
(\ref{mu-p}) we observe \be \Lambda_{dS}=2 H^2 \Big(1+2C
e^{-C}+\frac{1}{C}\Big)
>0\,. \label{ccds}\ee However, apparently the magnitude is not small
due a large $H$ (small $g^*$) around a $c=1$ fixed point, though
along the world sheet RG the system flows away from such a fixed
point to a supercritical regime and we do not know eventually to
what value of $H$ (or $g$) it settles to. Nevertheless, as we will
see in the last section, nonlocal processes like emission of baby
universes self-tunes $\Lambda_{dS}$ to a small positive value.


\section{The baby universes}
\setcounter{equation}{0}

Let us now discuss the nonlinear part of the WdW constraint that
connects the different topological sectors in the superspace. The
cosmological processes of topology change, such as creation of a
universe from nothing, or more generally loop splitting due to
reconnection of intersecting strings giving birth to baby universe
\cite{scole} or evolution of cosmic strings \cite{vs}, connect
different topological sectors of superspace (each such superspace
sector being characterized by all metrics having the same topology).
This suggests that the Wheeler-de Witt operator could be modified by
adding a local
 operator that has matrix elements between different
superspace sectors. As suggested by Vilenkin \cite{v-aqc}, this can
be schematically written as

\be \mc{H} \psi_{n}(l)+ \sum_{n' \ne n} ~\int [dl'] \Delta_{nn'}
(l,l') \psi_{n'}(l)=0 \,,\label{modifiedwdw}\ee where $l$ is the
superspace variable (here the length of the loop) and $n$ labels the
topological sector (here the number of disconnected loops or the
occupation number of the closed strings). Considering topology
change to be a local process, it can be assumed that
$\Delta_{nn'}(l,l')$ vanish unless $l$ and $l'$ are obtained from
each other by changing topological relations at a single point, {\it
i.e.} $\Delta_{nn'}(l, l')\ne 0$ for $n'=n \pm 1$. Knowing the
functional form of $\Delta_{nn'}(l, l')$ needs to deal with string
interaction vertices at the full quantum level. Even a qualitative
knowledge of its functional form in the $2D$ case will be immensely
helpful in understanding the topology changing amplitudes in the
higher dimensions.


\subsection{Nonlinear part of modified WdW equation}

Here we will derive the emergence of the topology changing
amplitudes from the nonlinearity in the WdW equation captured by
large $N$ RG . It is argued in \cite{cts1} that such a nonlinearity
in the WdW equation is essential in the matrix model description of
$2D$ gravity. However, there seems to be no control over the numbers
and sizes of the baby universes emitted or absorbed. Here we see
that the world sheet RG is very special in this respect that it
gives rise to baby universes of vanishing size with amplitudes
suppressed as the higher order corrections in $O(1/N)$ to WdW
equation. This implies that such processes are suppressed at large
scale and at weak string coupling. The suppression of the amplitudes
comes as a virtue of the vanishing size of the baby universes. It
will be extremely interesting to understand the underlying phenomena
that enables the world sheet RG in the large $N$ language to control
the size of the baby universes.

Let us recall (\ref{fullgenWDW}) and consider the multi-trace or
wormhole terms represented by the ellipses in the equation. In
matrix quantum mechanics they are redundant operators. Their effects
are lower by relative orders of $O(1/N)$ compared to the leading
contribution. To capture their corrections to the single topology
part of the WdW equation, let us rewrite the expansion $\tilde
\Sigma^{-1}$ from (\ref{SigmaInvFourier}) as

\ba && \Big \langle \tilde \Sigma^{-1} \mbox{Tr} \frac{e^{l
\phi(\hat x)}}{N} \Big \rangle_{\eta_i}\sim \Big
\langle\mbox{Tr}\frac {e^{l \phi(\hat x)}}{N} \Big
\rangle_{\eta_i}\nonumber\\&& +\Big \langle \Big[\int dt \int dl'
\delta(l') \Big(g~F_{g1} \frac{\partial}{\partial l'}+ g^2~
F_{g2}\frac{\partial^2}{\partial l'^2} + g^3~
F_{g3}\frac{\partial^3}{\partial l'^3}+...\Big) \mbox{Tr} e^{l'
\phi(t)}\nonumber\\&&+ \int dl' dl''~ \delta(l')\delta(l'')
\Big\{\int dt' dt''\Big (g^2~F_{g1}^2 \frac{\partial}{\partial l'}
\frac{\partial}{\partial l''}+2 g^3 F_{g1} F_{g2}
\frac{\partial}{\partial l'} \frac{\partial^2}{\partial
l''^2}+...\Big)\nonumber\\&&-\int dt' g^2~\hat F_{g2}\frac{1}{l'l''}
\frac{\partial^2 }{\partial t'^2}\Big\} \mbox{Tr} e^{l'
\phi(t')}\mbox{Tr} e^{l'' \phi(t'')}\nonumber\\&&+... \Big]
\mbox{Tr}\frac {e^{l \phi(\hat x)}}{N}\Big\rangle_{\eta_i}\,.
\label{expn-sigmainv}\ea Here $F_{gi}$-s are hyperbolic functions
given by (\ref{defhyperbolic}). Using the expansion in
(\ref{expn-sigmainv}) into (\ref{anomaly}) and recalling
(\ref{cso}), the inclusion of the topology changing amplitudes in
the beyond minisuperspace WdW can be arranged as

\be \mc{H}\psi_1(l)+ \sum_{n \ne 0,-1} \prod_{i,j=1}^{n}\int dl'_j~
\int dt_i \Big[\Delta_{1,n+1}(l;l'_j) \psi_{n+1}(l;l'_j,t_i)+ \tilde
\Delta_{1,n+2}(l; l'_j,t_i)\psi_{n+2}(l;l'_j, t_i)\Big]=0\,,
\label{topochange}\ee where $\mc{H}$ represents the WdW operator
acting on a given topological sector. Let us call this part single
topology WdW operator. Here $\mc{H}$ acts on wave function
$\psi_1(l)$ corresponding to a single loop and is given by
(\ref{generalWDW}) \be
\mc{H}=\Big(\frac{1}{l^2}-\frac{1}{l}\frac{\partial}{\partial
l}\Big)\Big(N\frac{\partial}{\partial
N}-\beta_{\eta_i}\frac{\partial}{\partial\eta_i}+1\Big)+A \cosh
\Big(\tilde l \frac{\partial}{\partial l}\Big)+ B \lambda_n
\sinh\Big(\tilde l \frac{\partial}{\partial l}\Big)\,. \ee
$\psi_n$-s represent $n$-point functions or the $n$-loop amplitudes

\be \psi_n(l_1,l_2,..l_n;t_1,t_2,..t_n)=\Big \langle \mbox{Tr}
\frac{e^{l_1 \phi(t_1)}}{N} \mbox{Tr}\frac{e^{l_2
\phi(t_2)}}{N}...\mbox{Tr}\frac{e^{l_n \phi(t_n)}}{N}\Big \rangle
\,. \ee To have a flavor about the nature of the matrix elements
between the different topological sectors, let us write down first
few matrix elements as differential operators

\ba &&\Delta_{1,2}(l;l')=N\delta(l') \sum_{m=1}\Big(g^m~F_{g m}
\frac{\partial^m}{\partial l'^m}\Big) \Big(A \cosh \Big(\tilde l
\frac{\partial}{\partial l}\Big)+ B \lambda_n \sinh\Big(\tilde l
\frac{\partial}{\partial
l}\Big)\Big)\,,\nonumber\\&&\Delta_{1,3}(l;l',l'')=N^2
\delta(l')\delta(l'') \sum_{m=1}\Big(g^m~F_{g m}
\frac{\partial^m}{\partial l'^m}\Big) \sum_{k=1}\Big(g^k~F_{g k}
\frac{\partial^k}{\partial l''^k}\Big) \nonumber\\&& \times \Big(A
\cosh \Big(\tilde l \frac{\partial}{\partial l}\Big)+ B \lambda_n
\sinh\Big(\tilde l \frac{\partial}{\partial l}\Big)\Big)\,,
\nonumber\\&&\tilde \Delta_{1,3}(l;l',l'',t')=-N^2 \delta(l')
\delta(l'')g^2~\hat F_{g2}\frac{1}{l'l''} \frac{\partial^2
}{\partial t'^2}\Big(A \cosh \Big(\tilde l \frac{\partial}{\partial
l}\Big)+ B \lambda_n \sinh\Big(\tilde l \frac{\partial}{\partial
l}\Big)\Big) \,. \ea They represent branching of a loop into $n$
loops out of which $(n-1)$ have vanishing radius. Thus the $(n-1)$
loops are emitted (absorbed) as `baby universes' from (by) the
original one. As differential operators the dependence on the length
of the mother loop and the daughter loops factorizes. However, their
interacting nature is unveiled by plugging in the scale dependence
of the $n$-loop wave function. It will be certainly interesting to
solve (\ref{topochange}) by some numeric ansatz for the scale
dependence of $\psi_n$ by and get the scale dependence of the
topology changing amplitudes.

To get a clear idea about the relative order of these multi-trace
(wormhole) contributions in $O(1/N)$, one can absorb the extra $N$
dependence in each amplitude by rescaling the length of the
vanishing loops as $l' \to \tilde l'=N l' \to 0 $. This arranges the
multi-trace or wormhole terms in relative strength of $O(1/N)$
compared to the single trace terms. For example, The correction due
to $\Delta_{1,2}$ begins from a relative $O(1/N)$ compared to the
single topology part of the WdW equation. Similarly, the correction
due to $\tilde \Delta_{1,3}$ is of relative $O(1/N^2)$ and the
correction due to $\Delta_{1,3}$ begins from a relative $O(1/N^2)$
compared to the single topology WdW equation. Note that, at a given
order the contribution from kinetic term of the matrix quantum
mechanics has much larger contribution than that coming from the
quadratic term or the cubic interaction vertices.

It will be extremely interesting to understand the physics in the
world sheet RG that controls the baby universes to have vanishing
size and hence being suppressed by $O(1/N)$. In the large $N$ RG,
from the point of view of the matrix quantum mechanics, this is very
closely attached to the matrix nature of $\phi(t)$ and hence to the
involvement of all the $N^2$ quantum mechanical degrees of freedom
in the dynamics. This indicates that perhaps a proper understanding
of the physics of these $N^2$ quantum mechanical degrees of freedom
in the strong coupling and at small scale is a key step towards real
cosmology.


\subsection{Self-tuning universe}

Here we will show how the sub-leading contribution to the linear
part of the modified WdW equation, namely the contribution coming
from the formation of one baby universe, self-tunes the cosmological
constant to a small positive value. As pointed out in
\cite{polyakov}, the mechanism of self tuning of the cosmological
constant by uncontrollable emission of baby universes \cite{scole}
has been studied extensively  in the literature. However, the main
problem is to understand the issues like the locality of the
coupling of the baby universes to the parent universe. As we have
discussed above, in large $N$ world sheet RG a controlled emission
of baby universes of vanishing size automatically emerges from the
nonlinear part of the modified WdW equation (\ref{topochange}).
These events have local coupling with the parent universe by the
virtue of the vanishing size of the baby universes. Below we will
derive the leading contribution of such event from the nonlinear
part of the modified WdW equation (\ref{topochange}) and its role in
self tuning the de Sitter cosmological constant (\ref{ccds}).

Considering the leading contribution from the emission of one baby
universe, the nonlinear modified WdW equation takes the form

\ba && \Big(\frac{1}{l^2}-\frac{1}{l}\frac{\partial}{\partial l}
\Big) \Big(N \frac{\partial}{\partial N}-\beta_{\eta_i}
\frac{\partial}{\partial \eta_i}+1 \Big) \psi_1(l,\hat x) +\Big(A
\cosh(\tilde l \frac{\partial}{\partial l})+B \lambda_n \sinh(\tilde
l \frac{\partial}{\partial l}) \Big)\nonumber\\&& \Big(\psi_1(l,\hat
x)+ \frac{1}{N}~\frac{ g \coth(\pi M R)}{2  M} \int dl' \delta (l')
\frac{\partial}{\partial l'} \int dt' \psi_2(l,l';\hat x,t')\Big) =0
\,.\ea However, integrating by parts the $O(1/N)$ term can be
simplified as

\be \int dl' \delta (l') \frac{\partial}{\partial l'}\int dt'
\psi_2(l,l';\hat x,t')=-\int dl' \Theta(l') \int dt' \psi_2
\,,\label{slcontribn}\ee where the boundary term $\delta(l') \int
dt' \psi_2(l,l';\hat x, t')$ vanishes $\psi_2$ being a continuous
function of $l'$ and $\Theta(l')$ represents the Heaviside step
function. Performing the $l'$ integration for $l' >0$
(\ref{slcontribn}) becomes proportional to

\be \int dl' \delta (l') \frac{\partial}{\partial l'}\int dt'
\psi_2(l,l';\hat x,t') \sim -e^{\infty} \int^t
\frac{dt'}{\lambda(t')} \psi_1\,, \ee where we have assumed all the
$N$ eigenvalues of $\phi(t)$ are same, given by $\lambda(t)$, and
the factor of order $e^{\infty}$ comes from the $l' \to \infty$ end
of the definite integral. Considering the periodic eigenvalue to be
$cos(\frac{n}{R} t)$, such a contribution will modify the de Sitter
cosmological constant (\ref{ccds}) by modifying the constant $A \to
A(1-\sigma)$ to be

\be \Lambda_{dS}'=2 H^2\Big(1-\frac{\mc{F}_1}{A(1-\sigma)}\Big)
=2H^2\Big(1+\frac{2Ce^{-C}+\frac{1}{C}}{1-\sigma}\Big)\,, \ee where
$\sigma$ is given by

\ba \sigma&=&\frac{1}{N} \frac{g R}{2 n M} \coth(\pi M R)\int^t
\frac{dt'}{\lambda(t')}~e^{\infty}\nonumber\\
&=& \frac{1}{N} \frac{g R}{2 n M} \coth(\pi M R) \sec(\frac{n}{R}t)
\tan(\frac{n}{R} t)~e^{\infty} \gg 1 \,.\ea Thus the de Sitter
cosmological constant becomes

\be \Lambda_{dS} \to \Lambda'_{dS}\sim 2 H^2
\Big(1-\frac{2Ce^{-C}+\frac{1}{C}}{\sigma}\Big)\,,\ee where it is
possible to get $\frac{2Ce^{-C}+\frac{1}{C}}{\sigma} \lesssim 1$ for
small $C$ as the numerator diverges in inverse power in $C$ whereas
the denominator diverges exponentially. Thus irrespective of the
magnitude of $H$, for small $C$ the sub-leading effect of the
emission of a single baby universe can achieve the desired self
tuning $\Lambda_{dS} \to \Lambda_{dS}' \gtrsim 0$. It will be
extremely interesting to capture the phenomena from some kind of
nonlocal world sheet action (see the world sheet methods in
\cite{abs, w-multi}).


\bigskip

\begin{flushleft}
{\bf Acknowledgments}
\end{flushleft}

We would like to thank Ofer Aharony, Micha Berkooz, Michael Douglas,
Gregory Gabadadze, Emil Martinec, Eva Silverstein for discussions
and useful comments. The work was partially supported by Feinberg
Fellowships, by the Israel-US Binational Science Foundation, the
European network HPRN-CT-2000-00122, the German-Israeli Foundation
for Scientific Research and Development, by the ISF Centers of
Excellence Program and Minerva.



\end{document}